\documentclass{aa} 

\usepackage{graphicx}
\usepackage{txfonts}
\usepackage{letltxmacro}
\usepackage{xcolor}
\usepackage{multirow}
\usepackage{booktabs}
\usepackage{amsmath}
\usepackage{csquotes}
\usepackage{placeins}
\usepackage{multicol}
\usepackage{float}
\usepackage{array}

\makeatletter
\newcommand*{\flushqueue}{\bgroup
  \let\mylist=\@deferlist
  \gdef\@deferlist{}\par
  \loop\@next\mybox\mylist{}{\let\mybox=\voidb@x}%
  \ifvoid\mybox
  \else
    \ifnum\count\mybox<64
      \begin{figure}[ht!]
        \unvbox\mybox
      \end{figure}\par
    \else
      \begin{table}[ht!]
        \unvbox\mybox
      \end{table}\par
    \fi
  \repeat
\egroup}
\makeatother

\newlength{\fsize}
\LetLtxMacro\oldselectfont\selectfont
\makeatletter
\DeclareRobustCommand{\selectfont}{\oldselectfont\setlength{\fsize}{\f@size pt}}
\makeatother

\newcommand\Msol{\ensuremath{\mathrm{M}_{\sun}}}
\newcommand\Zsol{\ensuremath{\mathrm{Z}_{\sun}}}
\newcommand\sfr{\ensuremath{\mathrm{SFR}}}
\newcommand\sfrir{\ensuremath{\mathrm{SFR}_{\mathrm{IR}}}}
\newcommand\sfruv{\ensuremath{\mathrm{SFR}_{\mathrm{UV}}}}
\newcommand\sfrratio{\ensuremath{\mathrm{SFR}_{20}/\mathrm{SFR}_{100}}}
\newcommand\sfrunits{\ensuremath{\Msol\,\mathrm{yr}^{-1}}}
\newcommand\ssfr{\ensuremath{\mathrm{sSFR}}}

\newcommand\micron{\ensuremath{\rm{\mu m}}}
\newcommand\logstellmass{\ensuremath{\log_{10}\left(M_*/\Msol\right)}}
\newcommand\loghalomass{\ensuremath{\log_{10}\left(M_{\rm{h}}/\Msol\right)}}
\newcommand\fracmass{\ensuremath{f_{M_*}}}
\newcommand\fracsfr{\ensuremath{f_{\sfr}}}
\newcommand\mufrac{\ensuremath{\mu_{1.5}}}
\newcommand\Halpha{\ensuremath{\mathrm{H}{\alpha}}}
\newcommand\halfmassradius{\ensuremath{r_{M_*/2}}}

\newcommand\agemw{\ensuremath{\rm{Age}_{\mathrm{MW}}}}
\newcommand\tlb{\ensuremath{t_{L}}}
\newcommand\kpc{\ensuremath{\mathrm{kpc}}}
\newcommand\pc{\ensuremath{\mathrm{pc}}}
\newcommand\zspec{\ensuremath{z_{\mathrm{spec}}}}
\newcommand\zphot{\ensuremath{z_{\mathrm{phot}}}}
\newcommand\njypix{\ensuremath{\mathrm{nJy\,pix^{-1}}}}
\newcommand\kms{\ensuremath{\mathrm{km\,s^{-1}}}}
\newcommand\Mpc{\ensuremath{\mathrm{Mpc}}}
\newcommand\vpec{\ensuremath{v_{\mathrm{pec}}}}
\newcommand\vtrans{\ensuremath{v_{\mathrm{trans}}}}
\newcommand\mdstar{\ensuremath{M_{\mathrm{d,star}}}}
\newcommand\mdgas{\ensuremath{M_{\mathrm{d,gas}}}}
\newcommand\rcl{\ensuremath{r_{\mathrm{cl}}}}
\newcommand\rcore{\ensuremath{r_{\mathrm{c}}}}
\newcommand\cgsdens{\ensuremath{\mathrm{g\,cm^{-3}}}}
\newcommand\densICM{\ensuremath{\rho_{\mathrm{ICM}}}}
\newcommand\anchorforce{\ensuremath{\Pi_{\rm{gal}}}}
\newcommand\densStar{\ensuremath{\Sigma_{\star}}}
\newcommand\densGas{\ensuremath{\Sigma_{\mathrm{gas}}}}

\newcommand\rtrunc{\ensuremath{r_t}}
\newcommand\disclenstar{\ensuremath{r_{\mathrm{d,\star}}}}
\newcommand\disclengas{\ensuremath{r_{\mathrm{d,gas}}}}
\newcommand\rprojrel{\ensuremath{r_{\mathrm{proj}}/r_{200}}}
\newcommand\deltavpecf{\ensuremath{\Delta v^{\rm{F0083}}_{\mathrm{pec}}}}
\newcommand\atid{\ensuremath{a_{\mathrm{tid}}}}
\newcommand\agal{\ensuremath{a_{\mathrm{gal}}}}
\newcommand\Mpert{\ensuremath{M_{\mathrm{pert}}}}
\newcommand\Mgal{\ensuremath{M_{\mathrm{gal}}}}

\usepackage[colorlinks=true,citecolor=blue]{hyperref}

\begin{document}

\title{Unveiling multiple physical processes on a cluster galaxy at $z=0.3$ using JWST}

\titlerunning{Multiple physical processes on a cluster galaxy at $z=0.3$}
\authorrunning{Peter J. Watson et al. } 

\author{
    Peter J. Watson\inst{\ref{INAF-OAPd}} $^{\href{https://orcid.org/0000-0003-3108-0624}{\includegraphics[scale=0.5]{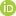}}}$\and 
    Benedetta Vulcani\inst{\ref{INAF-OAPd}} $^{\href{https://orcid.org/0000-0003-0980-1499}{\includegraphics[scale=0.5]{orcid.jpg}}}$\and
    Ariel Werle\inst{\ref{INAF-OAPd}} $^{\href{https://orcid.org/0000-0002-4382-8081}{\includegraphics[scale=0.5]{orcid.jpg}}}$\and
    Bianca Poggianti\inst{\ref{INAF-OAPd}} $^{\href{https://orcid.org/0000-0001-8751-8360}{\includegraphics[scale=0.5]{orcid.jpg}}}$\and
    Marco Gullieuszik\inst{\ref{INAF-OAPd}} $^{\href{https://orcid.org/0000-0002-7296-9780}{\includegraphics[scale=0.5]{orcid.jpg}}}$\and
    Michele Trenti\inst{\ref{UoMelbourne},\ref{ASTRO3D}} $^{\href{https://orcid.org/0000-0001-9391-305X}{\includegraphics[scale=0.5]{orcid.jpg}}}$\and
    Xin Wang\inst{\ref{SASS},\ref{NAO},\ref{IFAA}} $^{\href{https://orcid.org/0000-0002-9373-3865}{\includegraphics[scale=0.5]{orcid.jpg}}}$\and
    Namrata Roy\inst{\ref{WHMIII}} $^{\href{https://orcid.org/0000-0002-4430-8846}{\includegraphics[scale=0.5]{orcid.jpg}}}$
}

\institute{
    INAF -- Osservatorio Astronomico di Padova, Vicolo Osservatorio 5, 35122 Padova, Italy\\\email{peter.watson@inaf.it}\label{INAF-OAPd}\and
    School of Physics, University of Melbourne, Parkville, Vic 3010, Australia\label{UoMelbourne}\and
    Australian Research Council Centre of Excellence for All-Sky Astrophysics in 3-Dimensions, Australia\label{ASTRO3D}\and
    School of Astronomy and Space Science, University of Chinese Academy of Sciences (UCAS), Beijing 100049, China\label{SASS}\and
    National Astronomical Observatories, Chinese Academy of Sciences, Beijing 100101, China\label{NAO}\and
    Institute for Frontiers in Astronomy and Astrophysics, Beijing Normal University, Beijing 102206, China\label{IFAA}\and
    Center for Astrophysical Sciences, William H. Miller III Department of Physics and Astronomy, Johns Hopkins University, Baltimore, MD, 21218\label{WHMIII}
}

\date{Received 23 September, 2024; accepted 4 June, 2025}

    \abstract {
    We present a study of a previously identified candidate jellyfish galaxy in the Abell 2744 cluster, F0083, which showed faint signs of a tidal interaction in archival imaging.
    We used publicly available PSF-matched deep photometric data from the Hubble and James Webb Space Telescopes to infer the spatially resolved  star formation history of this galaxy.
    F0083 shows clear signs of ram-pressure stripping (RPS), with a recently enhanced star formation rate (\sfr) orientated towards the south-west quadrant of the stellar disc.
    The stellar mass surface density is heavily asymmetric, with a variation of nearly 1\,dex between the western spiral arm and the postulated tidal feature.
    This feature appears to contain a high proportion of older stars, ruling out RPS as the cause of this `unwinding'.
    We identified two potential interaction candidates, 28171 and 26055, with masses $\logstellmass=8.56\pm0.06$ and $\logstellmass=9.24\pm0.09$, respectively, and projected separations of 31\,\kpc\ and 46\,\kpc.
    The star formation history (SFH) of the tidal feature in F0083 indicates a steep change in \sfr\ at lookback times $\tlb\lesssim1\,$Gyr, consistent with a burst in the SFH of 26055.
    The most probable formation scenario of F0083 thus indicates a significant tidal interaction, followed by RPS as the combined system approaches pericentre passage.
    Our results demonstrate that by using photometric data we are able to distinguish between these consecutive processes, and represent the first observational analysis of the contributions of each process at this redshift.
  }


   \maketitle
%

\section{Introduction} \label{sec:intro}

\nolinenumbers
It has long been known that one of the primary driving forces in galaxy evolution is the surrounding environment \citep{oemler_systematic_1974,dressler_catalog_1980,larson_evolution_1980,cortese_dawes_2021}.
Compared to their counterparts in the field, galaxies within dense environments such as clusters have substantially different properties on average, often appearing redder, and comprising a higher fraction of ellipticals and fewer spirals \citep{dressler_catalog_1980}.
Whilst the more isolated field galaxies may only evolve through internal processes, the evolution of galaxies in clusters is heavily affected by external physical processes.
These can be divided into two main subclasses: gravitational and hydrodynamical.

Gravitational processes include all interactions between galaxies, such as tidal interactions and galaxy harassment \citep{merritt_relaxation_1983,byrd_tidal_1990,moore_galaxy_1996}.
These processes can affect the distribution of stellar mass within a galaxy, leading to extended tidal streams containing older stars, as well as inducing spiral structure, either symmetric or asymmetric depending on the relative interaction velocity \citep{wei_spatially-resolved_2021}.
They can disrupt the accretion and cooling of gas from the intergalactic medium (IGM) or compress the interstellar medium (ISM) in tidal features, leading to enhanced star formation rates (\sfr s) \citep{renaud_starbursts_2014}.
By contrast, hydrodynamic processes encompass all interactions between the dense plasma within the cluster halo, known as the intracluster medium (ICM), and the ISM of a galaxy moving at a range of velocities.
These interactions include ram-pressure stripping \citep[RPS;][]{gunn_infall_1972,takeda_ram_1984}, thermal evaporation \citep{cowie_thermal_1977}, starvation \citep{larson_evolution_1980,balogh_origin_2000}, and viscous stripping \citep{nulsen_transport_1982}.
These processes, in contrast to the gravitational processes, are able to directly remove the gas from the galaxy (or prevent it from accreting) without affecting the existing stellar content, usually resulting in a rapid cessation of star formation.

Ram-pressure stripping stands out amongst the hydrodynamical processes.
Whilst it is one of the most efficient methods of removing gas, depending on the orientation and relative velocity of the host galaxy with respect to the ICM \citep{schulz_multi_2001,vollmer_ram_2001,roediger_ram_2006,akerman_how_2023}, it does not initially prevent star formation.
Due to the progression of the galaxy through the cluster medium, the stripped gas can be observed in various phases as a tail.
This gas frequently cools into star-forming knots (but see \citealt{chung_virgo_2007,pedrini_muse_2022} for examples of galaxies where this does not happen), leading to extended tails that are visible across almost the entire electromagnetic spectrum, from X-ray to radio continuum \citep[e.g.][]{wang_x-raying_2004,sun_spectacular_2010,gullieuszik_uv_2023,moretti_observing_2022,gavazzi_radio_1995,scott_two_2012,roberts_lotss_2021,serra_meerkat_2024}.
Some of the most spectacular examples of RPS are known as jellyfish galaxies, so named for the peculiar morphology of these knots and filaments trailing the stellar disc \citep{smith_ultraviolet_2010,poggianti_gasp_2017}.

Due to the variety of interplaying physical processes, it remains difficult to obtain a full understanding of galactic evolution within high-density environments.
Some observations indicate that RPS is almost ubiquitous amongst blue galaxies in relatively local clusters \citep{boselli_virgo_2018,vulcani_relevance_2022}, and that the fraction observed is heavily dependent on the selection criteria.
On the other hand, galaxy mergers and other low-velocity interactions are thought to be rare in high-mass clusters, due to the high velocity dispersion of the system, which favours repeated high-velocity encounters \citep[harassment][]{moore_galaxy_1996}.

The identification of both gravitational and hydrodynamical processes processes in galaxies is not trivial, and typically requires deep observations across a wide range of the electromagnetic spectrum.
Examples have been found in the majority of local clusters where \ion{H}{I} or CO diagnostics can be utilised in addition to optical data, including Virgo \citep{vollmer_ngc_2003,vollmer_new_2005,boselli_virgo_2018}, and Leo \citep{gavazzi_75_2001,scott_two_2012,fossati_muse_2019}.
More recently, \cite{serra_meerkat_2024}, investigating the \ion{H}{I} distribution of NGC\,1427A ($\zspec=0.007$) in the Fornax cluster, have concluded that its properties cannot be explained by RPS alone, and are suggestive of a previous high-speed merger.
In the optical regime, using the Multi-Unit Spectroscopic Explorer (MUSE), \cite{fritz_gasp_2017} detail the jellyfish galaxy JO36 ($\zspec=0.041$), which appears to have been partially stripped by interactions with the ICM before a gravitational interaction distorted the remaining gas and stellar disc.
Looking at non-cluster environments, \cite{vulcani_gasp_2021} find only a single galaxy in which multiple processes can be identified using MUSE, from a sample of 27 asymmetric field galaxies.
The interaction between these gravitational and hydrodynamical processes is much less explored at earlier epochs, where the difficulties in identifying both processes increase (see \citealt{cortese_strong_2007} for an example at $z\sim0.2$, where the tidal interaction is ascribed to the cluster potential rather than another galaxy).
We aim to demonstrate that by using only multi-band photometry to derive spatially resolved star formation histories (SFHs), it is possible to isolate both the location and duration of these processes beyond the Local Universe.

The target environment for this study is the Abell 2744 galaxy cluster,  hereafter A2744 ($\rm{R.A.}=3.58641$, $\rm{Dec.}=-30.39997$; $\zspec=0.3064$), with a virial mass of $7.4\times10^{15}\,\Msol$ and radius $r_{200}=2\,\Mpc$ \citep{jauzac_extraordinary_2016}.
This is an extremely well-observed system, and is a highly complex environment, which gives rise to the nickname Pandora's Cluster \citep{merten_creation_2011}.
It has long been known to have significant merger activity, based on the identification of an extended radio halo and radio relic \citep{giovannini_radio_1999,govoni_comparison_2001,govoni_radio_2001}.
More recent observations have identified at least four radio relics in the vicinity of the cluster, up to 1\,Mpc away from the centre \citep{pearce_vla_2017,rajpurohit_dissecting_2021,golovich_merging_2019}, and X-ray data indicate multiple peaks, shock fronts, and cold fronts \citep{kempner_chandra_2004,owers_dissection_2011,merten_creation_2011,jauzac_extraordinary_2016,chadayammuri_closing_2024}.
The cluster itself contains no fewer than four interacting subclusters, with some disagreement on the precise location, mass, and membership of these structures, depending on whether they are derived from velocity distributions \citep{boschin_internal_2006,rawle_star_2014,vulcani_early_2023} or gravitational lensing maps \citep{merten_creation_2011,medezinski_frontier_2016,kokorev_alma_2022,bergamini_glass-jwst_2023,cha_precision_2024}.

The dynamical state and multiple components of A2744 make it extremely difficult to find any consensus on its formation history.
Thus far, the two most robust claims are that one merger has occurred near the plane of the sky, in the north-south direction, and that another has occurred in the south-east of the cluster, with a large line-of-sight velocity component.
Whilst there are additional hypotheses relating to substructures in the west or north-west of the cluster \citep{kempner_chandra_2004, merten_creation_2011,medezinski_frontier_2016}, they are not relevant to our analysis, and are not considered further.

The N-S merger is often referred to using the substructure nomenclature of \cite{owers_dissection_2011}, as the northern (major remnant) core (NC) and the central tidal debris (CTD).
The third major substructure, the southern minor remnant core (SMRC), has a projection coincident with the CTD, but is offset in velocity space by $\sim5000\kms$, and may not even be gravitationally bound to the remainder of the cluster \citep{chadayammuri_closing_2024}.
These names, however, may be misleading. 
More recent analyses converge on a scenario where the majority of mass is contained within the southern core of the cluster, the CTD, and both the NC and SMRC have passed through this region within the last 1 Gyr \citep{mahler_strong-lensing_2018,furtak_uncovering_2023,bergamini_glass-jwst_2023,cha_precision_2024}.

\begin{figure}
    \centering
    \includegraphics[width=\hsize]{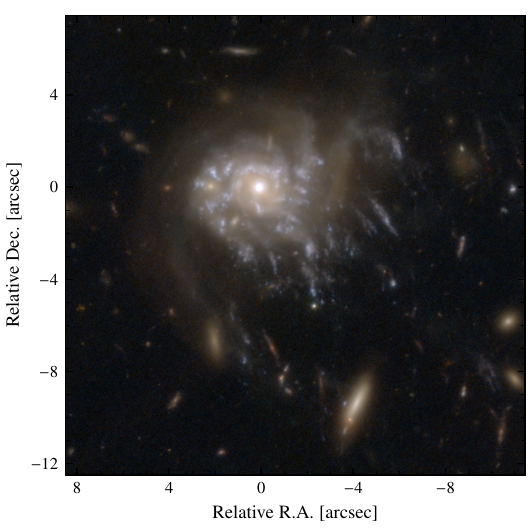}
    \caption{
        Colour image of the galaxy F0083, created by combining publicly available imaging from the Hubble and James Webb Space Telescopes (HST and JWST) at their native resolution, in the following bands: F435W, F606W, F070W, F814W, F090W, F115W, F150W, and F200W.
    }
    \label{fig:F0083_full_colour_image}
\end{figure}

The main subject of this paper is a large ($\approx12\arcsec\times12\arcsec$, or $55\times55\,\kpc$) spiral galaxy, shown in colour in Fig.~\ref{fig:F0083_full_colour_image}. 
It is located on the south-eastern side of Abell 2744 ($\rm{R.A.}=3.61066$, $\rm{Dec.}=-30.39560$), close to the cluster core at $\rprojrel=0.18$ and at a redshift $z=0.3033$ ($\vpec=-712\kms$), and is consistent with being a member of either the CTD substructure or the general cluster environment.
This unique galaxy, hereafter referred to as F0083 after \cite{owers_shocking_2012}, has previously been identified as a potential jellyfish or RPS galaxy, with a possible tidal feature (see below for a comprehensive overview of the published literature).
However, the studies thus far have not conclusively established that either of these mechanisms affect F0083, and whether there are any alternative explanations for its formation and evolution.

For example, there are many instances of more local galaxies with highly disturbed morphologies and elevated SFRs, resulting from mergers rather than interactions within a cluster environment (the Whirlpool, Antennae, and Mice galaxies being amongst the most well known).
Whilst F0083 is located within the cluster, it is possible that interactions with other galaxies have dominated the history of this object, resulting in the observed morphology.
On the other hand, it has recently been suggested that one of the effects of RPS may be to unwind the spiral arms of galaxies, due to the differential pressure exerted as a result of the disc rotation \citep{bellhouse_gasp_2017, bellhouse_gasp_2021,vulcani_relevance_2022}.
These different mechanisms should each leave a distinctive mark on the SFH of the galaxy (and any interacting neighbours), and it is this that we aim to exploit, through spatially resolved SED fitting to deep photometric observations.

F0083 has been observed previously, in a small number of studies, and we discuss these briefly below.
The first is \cite{owers_shocking_2012}, from which we adopt their assigned name, who identified this object as a potential jellyfish galaxy using HST/ACS imaging in the F435W, F606W, and F814W bands.
They identified the presence of a low-luminosity active galactic nucleus (AGN, $L_{\rm{X}}\sim 10^{41}$\,erg\,s$^{-1}$) based on Chandra X-ray observations of the cluster field, and through double Gaussian fits to single fibre spectroscopy on the galaxy centre, using AAOmega on the Anglo-Australian Telescope \citep{owers_dissection_2011}.
Intriguingly, they also suggested the presence of a tidal feature, connecting to a faint galaxy on the eastern side, indicating an interaction.
However, due to the data available at the time, it is not possible to comment further on this here.

\cite{rawle_star_2014} investigated a number of galaxies in the A2744 field, amongst them F0083.
Using GALEX NUV and SPITZER/IRAC observations, they estimated the UV and IR \sfr s as $\sfruv=26.2\pm0.9\,\sfrunits$ and $\sfrir=8.0\pm1.0\,\sfrunits$ respectively, accounting for the AGN luminosity in the IR regime.
From the \sfruv/\sfrir\ ratio, they suggested that F0083 is likely starbursting, and has been observed near peak star formation efficiency as the dusty gas reservoir has been almost completely exhausted.
They also estimated the total stellar mass using IRAC 3.6 and 4.5 \micron\ fluxes, using the relation from \cite{eskew_converting_2012}, giving a result of $\logstellmass\approx10.91$, for a \cite{kroupa_variation_2001} initial mass function (IMF).
\cite{rawle_star_2014} also posited that an infall scenario was unlikely for this galaxy, as the orientation of the visible blue tail is not aligned with the radial line of the cluster core, and that it had been stripped due to shock fronts from mergers of the cluster cores themselves, although there is no clear consensus on the formation history of this cluster \citep{merten_creation_2011,golovich_merging_2019,rajpurohit_dissecting_2021,cha_precision_2024}, and such mechanisms have rarely been explored, even in simulations \citep{ruggiero_galaxy_2019}.

More recently, \cite{lee_enhanced_2022} and \cite{lee_gmosifu_2022} observed F0083 in part with the GMOS IFU, over the range 7820--9260\,\AA.
Whilst this only covered a limited field of view ($5\arcsec\times7\arcsec$) encompassing the central component and a portion of the blue tail, they estimated a total $\sfr=22.0\,\sfrunits$ in the region they observed (accounting for AGN emission), with $\approx35\%$ of star formation occurring in the tail.
However, it is worth noting that the \Halpha-derived \sfr s from \cite{lee_gmosifu_2022} trace more recent star formation than either the \sfruv\ or \sfrir\ used by \cite{rawle_star_2014}, and are not directly comparable.
They highlight that the galaxy shows clear signs of a rotating gas disc, and that the \Halpha\ flux contours are highly asymmetric, being extended towards the eastern region.

The central hypothesis of this paper is that F0083 represents another example of a galaxy undergoing RPS, following an interaction with a nearby galaxy producing a tidal stream.
In Section~\ref{sec:method} we present the data and the methods utilised to extract the necessary measurements. We show the results for F0083 in Section~\ref{sec:results}.
In Section~\ref{sec:discussion} we discuss and interpret these results, comparing them against the existing literature and simulations.
We also identify nearby galaxies that may have interacted, and contrast their derived SFHs against F0083.
Finally, in Section~\ref{sec:conclusions}, we summarise the salient points of this work, and discuss the implications and potential for further study.
Throughout this paper, we assume a \cite{kroupa_variation_2001} IMF and a solar metallicity $\Zsol=0.0142$ \citep{asplund_chemical_2009}.
Coordinates are given in the International Celestial Reference System (ICRS), and we assume a standard $\Lambda$ cold dark matter cosmology, with $\Omega_{\rm{M}}=0.3$, $\Omega_{\Lambda}=0.7$, and $h=0.7$.

\section{Method} \label{sec:method}

Whilst F0083 is the primary target, we  also considered other nearby galaxies.
The data and methods detailed here apply to all such galaxies throughout this study.

\subsection{Available data} \label{sec:method_data_reduction}

The data used in this study comprise all publicly available imaging using NIRCam and NIRISS on JWST, as well as WFC3/UVIS, WFC3/IR, and ACS/WFC on HST.
The main contributions to the JWST imaging were GLASS-JWST ERS \citep{treu_glass-jwst_2022,paris_glass-jwst_2023, merlin_astrodeep-jwst_2024}, UNCOVER \citep{bezanson_jwst_2024,weaver_uncover_2024}, and MegaScience \citep{suess_medium_2024}, with a full list of all contributing programmes in Appendix~\ref{app:programmes}.
We used here the drizzled mosaics made public through the most recent MegaScience data release \citep{suess_medium_2024},\footnote{These data releases are presently available at \url{https://jwst-uncover.github.io/DR3.html}} which combined all of these existing observations.
The mosaics were reduced using the ``jwst\_0995.pmap'' calibration reference file, and aligned and co-added using the Grism Redshift \& Line pipeline \citep[\textsc{grizli},][]{brammer_grizli_2019}.
All mosaics are normalised to have flux intensity units of 10\,\njypix, corresponding to an AB magnitude zeropoint of 28.9, and are sampled onto a grid with a scale of 0\farcs 04 per pixel.
At the redshift of the A2744 cluster ($z=0.3064$), this corresponds to a physical scale of 180\,\pc\ per pixel.

\subsection{PSF matching} \label{sec:method_psf_matching}
For accurate spatially resolved photometry, it is necessary to ensure that all measurements are made on images with a similar point-spread function (PSF).
Given the large variation in the size and shape of the PSF, from the UVIS filters on HST/WFC3 to the long-wavelength (LW) filters on JWST/NIRCam, this requires deriving convolution kernels to match a target PSF.
For simplicity, we followed the lead of \cite{suess_medium_2024}, and matched to the NIRCam F444W filter.
The MegaScience public data release includes empirical PSFs and convolution kernels for 27 of the 33 filters with coverage in the regions of interest.
We therefore used these PSFs and kernels where available, and derived our own PSFs for the remaining filters, with the procedure detailed in Appendix~\ref{app:psf_matching}.

\begin{figure}
    \centering
    \includegraphics[width=\hsize]{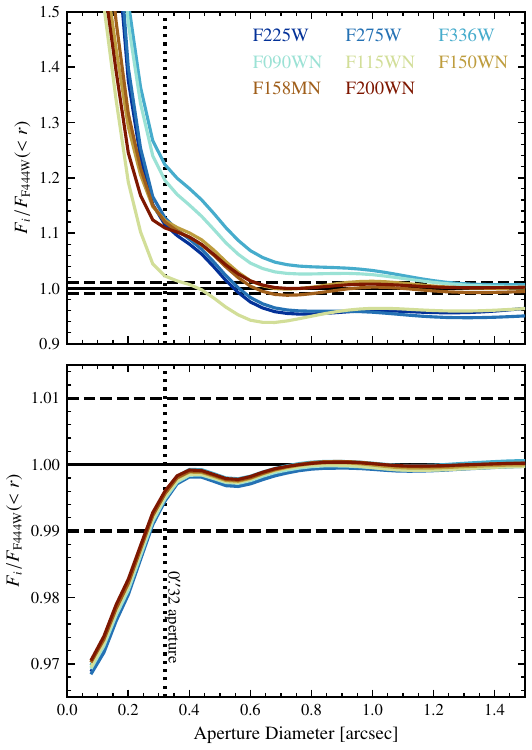}
    \caption{
        PSF curve of growth for each filter in our analysis, before (top) and after (bottom) matching to the F444W PSF.
        The curves are shown as the encircled flux relative to the F444W growth curve, where unity would imply an exact match.
        The solid black line indicates an exact match, the dashed lines the $\pm1\%$ deviations from this, and the dotted line the location of a 0\farcs32 aperture, for ease of comparison with the growth curves shown in \cite{weaver_uncover_2024} and \cite{suess_medium_2024}.
    }
    \label{fig:psf_cog_matching}
\end{figure}

In Fig.~\ref{fig:psf_cog_matching}, we compare the encircled energy of our matched PSFs against the target, as a function of the aperture diameter.
Comparing against the results shown in Fig.~6 of \cite{weaver_uncover_2024} and Fig.~3 of \cite{suess_medium_2024}, our results are of a similar quality to the existing PSF-matching kernels.
The maximum deviation from the target PSF is $\approx3\%$, at an aperture diameter of 0\farcs06 (1.5 pixels), consistent with the residual pattern shown in Fig.~\ref{fig:psf_kernel_comparison}e.
Some variation is unavoidable, given the vastly different shape and substructure of the PSFs, but we highlight that at apertures larger than $\approx0\farcs27$ (1.22\,\kpc\ at $z=0.3064$), the encircled flux in all filters is consistent within $1\%$ of the F444W target.

\subsection{Voronoi binning}
\label{sec:method_voronoi_binning}

To maintain a sufficient signal/noise (S/N) when deriving the spatially resolved properties, it is necessary to bin the data in some form.
For our science case, we aimed to satisfy two criteria:
\begin{itemize}
    \item The S/N in each bin must be sufficient to reliably constrain the SFH, given the non-parametric form detailed in Section~\ref{sec:method_sed_fitting}.
    \item The bin size must be large enough to avoid sampling at scales where the PSF convolution is highly uncertain.
\end{itemize}
On the latter point, we required a minimum bin radius of 0\farcs27, as this was the minimum diameter for which the encircled energy of every matched PSF deviated by less than $1\%$ from the target (see Section~\ref{sec:method_psf_matching}).
We therefore adopted a modified version of the Weighted Voronoi Tesselation technique, developed by \cite{cappellari_adaptive_2003} and \cite{diehl_adaptive_2006}, in which bin shapes tend towards close-packed hexagons.
Consequently, during the initial step of the algorithm, we restricted the progression such that a new bin could not be initiated until the current bin had accreted at least 30 pixels, the area corresponding to a hexagon with circumcircular diameter 0\farcs27.
To calculate the S/N, we used the flux in the NIRCam F150W filter as the signal, and for the noise the square root of the full variance array in the same filter, including contributions from the read noise, sky background, and the Poisson noise from individual sources. 
We targeted a $\rm{S/N}=100$, resulting in 1637 discrete bins.

\subsection{Redshifts} \label{sec:method_redshift}

Throughout this paper, for cluster galaxies other than F0083 we make use of redshifts from the MegaScience data release catalogue \citep{suess_medium_2024}.
These photometric redshifts, \zphot, were derived through SED fitting to the integrated flux in all available photometric bands, using the SED fitting tool \textsc{prospector} \citep{johnson_stellar_2021}, and the physical model of \cite{wang_uncover_2024}. 
We refer the reader to the aforementioned papers for further details.
For a small number of galaxies in the A2744 field, there exist spectroscopic redshifts, \zspec.
We cross-matched the locations of galaxies in the MegaScience catalogue against the spectroscopic catalogue compiled by \cite{bergamini_glass-jwst_2023}, the NIRSpec catalogue of \cite{mascia_glass-jwst_2024}, and the NIRISS catalogue of \cite{watson_glass-jwst_2025}.
Where available, we make use of these measurements instead of the maximum-likelihood photometric redshifts.

\subsection{SED fitting} \label{sec:method_sed_fitting}

\begin{figure*}
    \centering
    \includegraphics[width=\textwidth]{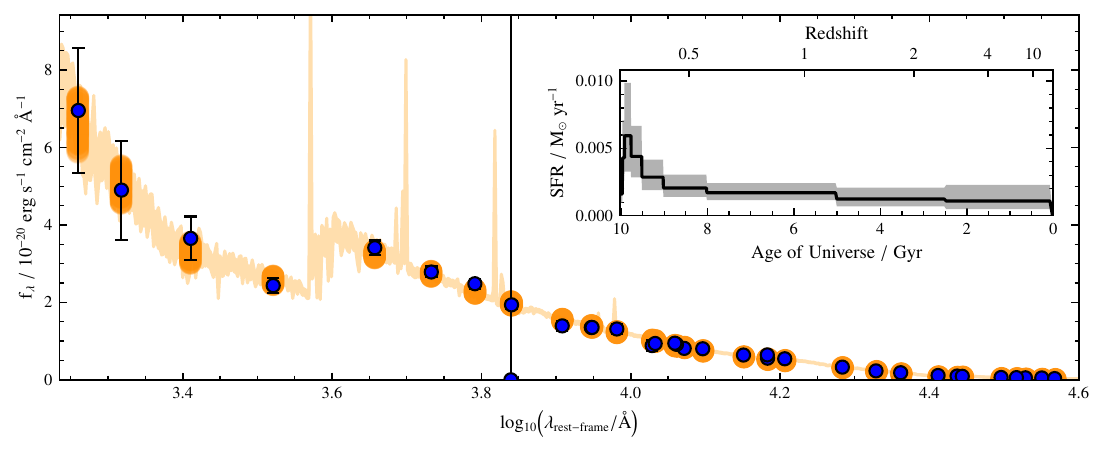}
    \caption{
        Example \textsc{bagpipes} fit to the photometry in a single Voronoi bin, plotted against the rest-frame wavelengths.
        The blue points indicate the measured flux at the pivot point of each filter, with their uncertainty (including the additional systematic component; see Appendix~\ref{app:photometric_uncertainties}).
        The shaded orange region indicates the 16${\rm{th}}$-84${\rm{th}}$ percentile of the model spectrum, and the orange points the corresponding predicted fluxes in each filter.
        In the inset, we show the derived SFH of this region as a function of both redshift and the age of the Universe, with the shaded region denoting the 16${\rm{th}}$-84${\rm{th}}$ percentile of the posterior PDF in each age bin.
    }
    \label{fig:bagpipes_example_fit}
\end{figure*}

We modelled the stellar populations of each galaxy by fitting the observed fluxes in all photometric bands with the SED fitting code, \textsc{bagpipes} \citep[Bayesian Analysis of Galaxies for Physical Inference and Parameter EStimation,][]{carnall_inferring_2018}.
For Voronoi-binned galaxies, the fluxes in each band were measured as the sum of the fluxes of all pixels within each Voronoi bin.
The uncertainties were measured similarly, albeit with an additional $5\%$ uncertainty term added in quadrature to account for systematic effects (see Appendix~\ref{app:photometric_uncertainties}).
\textsc{bagpipes} uses the \cite{bruzual_stellar_2003} spectral library,\footnote{\url{https://www.bruzual.org/~gbruzual/bc03/Updated_version_2016/}} updated with the 2016 version of the MILES library of empirical spectra  \citep{falcon-barroso_updated_2011}.
This spatially resolved SED fitting relies on the assumption that each spatial element is independent.
Whilst this is certainly not true in general, the minimum feature size probed in this study ($\geq1\kpc$) ensures that the energy transfer from adjacent regions is negligible \citep{boquien_measuring_2015,smith_panchromatic_2018}.

Whilst spatially resolved photometric SED fitting is not a novel concept \citep[][]{conti_star_2003,zibetti_resolved_2009,sorba_missing_2015,sorba_spatially_2018}, much of the previous literature has limited the fit to simple parametric forms for the SFH \citep{wuyts_smoother_2012,abdurrouf_introducing_2021}, which can often fail to recover realistic SFHs in scenarios such as mergers \citep{smith_deriving_2015}.
Here, we adopted a non-parametric form for the SFH, following the method of \cite{leja_how_2019}.
We fitted for a constant \sfr\ in multiple bins of the lookback time $t_L$, where the bin edges were set at 0, 0.02, 0.05, 0.1, 0.25, 0.5, 1, 2, 5, 7.5, and 10\,Gyr.
The exact edge of the final bin was determined by the age of the Universe at the redshift of the fitted galaxy, i.e. $\approx10$\,Gyr for F0083 at $z=0.3033$.
Dust attenuation was described by the Milky Way extinction curve of \cite{cardelli_relationship_1989}, with $R_V=3.1$.
We allowed for the attenuation $A_V$ to vary between zero and two magnitudes, and for $\eta$, the multiplicative factor on $A_V$ for stars in birth clouds, to vary as $0.5\leq\eta\leq3$, both with uniform priors.
Emission lines and nebular continuum emission were included in the fit for stellar populations as old as 20\,Myr, via a method based on \cite{byler_nebular_2017}, using the \textsc{cloudy} photoionisation code \citep{ferland_2017_2017}.
The ionisation parameter $\log U$ was allowed to vary between $-3.5$ and $-2$.
The stellar metallicity was assumed to be the same as the gas-phase metallicity, and was allowed to vary between 0.01 and 2.5\,\Zsol, following a Gaussian prior with mean $\mu=\Zsol$ and standard deviation $\sigma=0.5\,\Zsol$.
The redshift was fixed to the galaxy redshift (see Section~\ref{sec:method_redshift}), as variations in velocity within a galaxy have a negligible effect on the broad-band photometry \citep{werle_history_2024}.

To sample the posterior distribution, \textsc{bagpipes} uses the \textsc{MultiNest} nested sampling algorithm \citep{feroz_importance_2019}, through the \textsc{PyMultiNest} interface \citep{buchner_x-ray_2014}.
For our data, we ran \textsc{bagpipes} with the default \textsc{MultiNest} sampling parameters, using 400 live points.
We show an example of the resulting photometric fit in Fig.~\ref{fig:bagpipes_example_fit}.
For bins in which the flux in one of the photometric bands was unusable (e.g. due to pixels masked during the mosaic creation), we set the flux to zero, and the uncertainty to $10^{30}$, which effectively discarded that filter from the fit.

\subsection{Contamination masking} \label{sec:method_contamination_masking}

We defined the extent of F0083 using \textsc{SExtractor} \citep{bertin_sextractor_1996}, by finding the connected region $3\sigma$ above the local sky background, measured using a stack of the JWST/NIRCam imaging in the F115W, F150W and F200W filters.
We opted to initially mask out the central seven bins for F0083, covering an approximate diameter of 0\farcs5 (2.2\,\kpc).
The reasons for doing so were two-fold.
Firstly, as F0083 hosts a low-luminosity AGN, there is a non-negligible contribution from this component to the photometry in bins near the centre of the galaxy.
The spectral energy distribution of AGN can appear extremely similar to the continuum shape of young stellar populations, due to the underlying non-thermal emission, and it is not possible to break this degeneracy using only broad-band photometry.
This can introduce significant biases in SED fitting codes such as \textsc{bagpipes}, leading to the inference of much younger and more massive stellar populations, than without the AGN contribution \citep{ciesla_constraining_2015,cardoso_impact_2017}.
Given the target F444W PSF, this masked region should contain $>85\%$ of the point-source flux.
We note, however, that the extended structure of the PSF (the so-called ``snowflake`` substructure, and the diffraction spikes) will still contribute a non-negligible amount of flux to the remainder of the bins (see Section~\ref{sec:results_sfh}, specifically Fig.~\ref{fig:F0083_regions_age_maps}).
Whilst this artefact may have some small impact on our results, we do not consider it significant enough to affect the overall conclusions.
The secondary factor is that the central pixels are saturated or show non-linear behaviour in a number of bands, notably the long-wavelength NIRCam bands, F277W, F356W, and F444W, in an aperture of $\approx0\farcs25$ diameter.
These bands are vital for deriving accurate estimates of older stellar populations, and hence the total stellar mass.
To avoid any potential bias in our conclusions, we prefer to mask this region entirely.

We also highlight that within the extent of F0083, the MegaScience data catalogue contains six other objects (see Fig.~\ref{fig:F0083_full_colour_image}).
Using the best-fit redshifts from Section~\ref{sec:method_redshift}, we discovered that only one of these sources was within the approximate range of the cluster, $0.29\leq z\leq0.32$ (see Section~\ref{sec:discuss_tidal_interaction} for further details).
Without dedicated spectroscopy, it is not possible to disentangle the flux contributed by any of these overlapping objects. 
Following our masking of the central AGN, we again opted for a conservative approach, and masked out the contaminated pixels (based on the \textsc{SExtractor}-derived segmentation map), visible in Fig.~\ref{fig:mass_map_radial_profile} as black regions.
As a result of these masked areas, we stress that some of the integrated properties of our galaxy should be considered a slight underestimate of the true values, specifically the stellar mass and \sfr.

\section{Results} \label{sec:results}

Whilst there are many derived quantites from our SED-fitting procedure, we focus here on the most relevant properties for our investigation that are well-constrained by the broad-band photometry.
In particular, we look at the distribution of stellar mass and the \sfr, both at the present epoch and as a function of lookback time \tlb.

\subsection{Stellar mass distribution} \label{sec:results_mass}

\begin{figure}
    \centering
    \includegraphics[width=\hsize]{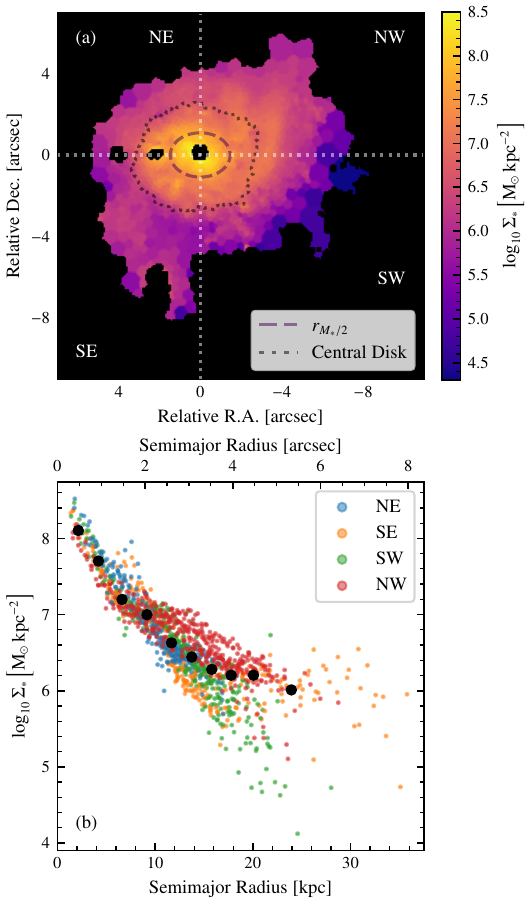}
    \caption{
        (a) Spatially resolved map of the stellar mass surface density for F0083, with all external objects masked out.
        We overlay the half-mass radius and the contour we define as encompassing the central disc in dashed and dotted lines, respectively.
        (b) Radial profile of the stellar mass surface density, as a function of the semi-major axis length.
        Points are coloured according to the quadrant in which they lie.
        We also show the median profile, averaged in ten radial bins.
    }
    \label{fig:mass_map_radial_profile}
\end{figure}

From our \textsc{bagpipes} fits, we obtain an estimate of the current stellar mass in each Voronoi bin.
Summing over all of the bins, we estimate the total stellar mass of F0083 to be $\logstellmass=10.24\pm0.03$.
Compared against other studies such as \cite{rawle_star_2014}, this appears substantially lower by $\sim0.6$\,dex, although there are a number of contributing factors for this.
Previous measures of the stellar mass have relied on the integrated flux in IRAC 3.6 and 4.5\,\micron\ bands, using the relation of \cite{eskew_converting_2012} to derive the stellar mass.
However, in this study, they caution that the mass estimates are significantly affected by even a modest fraction of young stars in the population,\footnote{An exact relation between the fraction of young stars and the estimated mass is not given, but the residuals are as high as 0.5\,dex.} which itself is almost certainly large given the morphology and colour of F0083.
Finally, in the same way that our measure is likely an underestimate due to our conservative masking procedure (see Section~\ref{sec:method_contamination_masking}), it is highly likely that other measures overestimate the IR flux due to overlapping sources.
Indeed, if we look at all bins for which the objects overlap in the original segmentation map, we instead find a mass of $\logstellmass=10.41\pm0.1$.
As such, we do not consider this a significant discrepancy.
We consider also the possibility of estimating the total mass from our measured values, without masking.
We interpolate over the masked regions using a linear spline (and extrapolate the profile in Fig.~\ref{fig:mass_map_radial_profile}b to the centre), to obtain an estimate of $\logstellmass=10.30\pm0.15$.
This is not a substantial enough difference to change any of our results or conclusions, and so we maintain the use of the direct measurement throughout this paper.

In Fig.~\ref{fig:mass_map_radial_profile}a, we present the 2D map of the stellar mass surface density of F0083.
We overlay the best-fit ellipse for the half-mass contour, enclosing half of the stellar mass, as a dashed line, giving a semi-major radius of $\halfmassradius=1.09\pm0.01\,\arcsec$ (or $4.89\,\kpc$), ellipticity $\epsilon=0.26$, and position angle $\theta=83.9$.
It is immediately clear that this galaxy has a disturbed morphology, with a significant degree of asymmetry in the mass distribution.
In Fig.~\ref{fig:mass_map_radial_profile}b, we highlight this by colouring the stellar mass surface density in each quadrant (with respect to the galaxy centre).
Whilst we see a monotonic decrease for bins in the north-eastern (NE) quadrant of F0083, there is a spike in stellar mass at a distance of $\approx8\,\kpc$ in the SE quadrant.
This corresponds to an extremely large star-forming region (see Section~\ref{sec:results_ssfr}), visible in the direct imaging in Fig.~\ref{fig:F0083_full_colour_image}.
In the SW quadrant, the stellar mass density appears to flatten at large radii ($>15\,\kpc$), due to the recent star formation in the tail, whereas the NW region briefly plateaus between $8<r<12\,\kpc$.
This plateau is due to the presence of a large spiral arm, and there is little to no evidence for any equivalently massive structure on the opposing side, with an offset of $\approx0.5\,$dex visible in Fig.~\ref{fig:mass_map_radial_profile}b.
There is however, a small increase in the mass surface density at an offset of $\sim5\farcs5$ from the centre towards the SE, in the location of the tidal feature reported by \cite{owers_shocking_2012}.
Whilst faint, this feature is also apparent as a yellow-red stream in the colour image in Fig.~\ref{fig:F0083_full_colour_image}.

Many other studies of jellyfish galaxies use some estimate of the extent of the stellar disc, or the galaxy main body, as a point of comparison, derived through a variety of methods -- e.g. manual masking \citep{lee_gmosifu_2022}, $g$-band isophotes \citep{moretti_observing_2022}, or the stellar continuum from spectral fitting \citep{gullieuszik_gasp_2020}.
Here, we opt to derive an estimate from the F606W filter (the closest equivalent to rest-frame $g$).
We adopt an ellipse with the same ellipticity and position angle as the half-mass contour (to prevent inclusion of the entire visible tail), and fit for the semi-major radius that matches the flux isophote $5\sigma$ above the sky background.
We refer to this as the ``central disc'', and show its extent in Fig.~\ref{fig:mass_map_radial_profile}.
This allows us to compare the SFHs in the exterior regions of F0083, whilst also providing a point of comparison to other studies.
For reference, this region encompasses $81\%$ of the total stellar mass of F0083 ($\logstellmass=10.15$, see Table~\ref{tab:F0083_properties}).

\begin{table*}
\caption{Properties of F0083 in different regions.}   
\label{tab:F0083_properties}    
\centering                       
\begin{tabular}{l c c c c c c c} 
\hline\hline                
Region & \logstellmass & \fracmass & \agemw\,[Gyr] & \mufrac & $\sfr_{100}$\,[\sfrunits] & \fracsfr & \sfrratio \\  
\hline    
No mask\tablefootmark{a} & $10.32\pm0.03$ & $120.1\%$ & $3.42$ & $39\%$ & $18.12\pm3.86$ & $125.5\%$ & 1.78 \\ 
Overlap Masked & $10.30\pm0.04$ & $113.8\%$ & $3.28$ & $41\%$ & $18.00\pm3.86$ & $124.7\%$ & 1.79 \\ 
Centre Masked\tablefootmark{b} & $10.24\pm0.03$ & $100.0\%$ & $3.24$ & $40\%$ & $14.44\pm0.42$ & $100.0\%$ & 1.31 \\ 
Central Disc\tablefootmark{d} & $10.15\pm0.03$ & $80.4\%$ & $3.07$ & $42\%$ & $11.79\pm0.41$ & $81.6\%$ & 1.36 \\ 
Exterior\tablefootmark{e} & $9.54\pm0.02$ & $19.6\%$ & $3.95$ & $32\%$ & $2.65\pm0.08$ & $18.4\%$ & 1.13 \\ 
   \hline
A & $9.10\pm0.02$ & $7.3\%$ & $4.44$ & $21\%$ & $0.46\pm0.03$ & $3.2\%$ & 0.53 \\ 
B & $8.75\pm0.02$ & $3.2\%$ & $3.47$ & $37\%$ & $0.24\pm0.02$ & $1.6\%$ & 0.73 \\ 
C & $8.79\pm0.03$ & $3.5\%$ & $3.86$ & $32\%$ & $0.24\pm0.02$ & $1.7\%$ & 0.83 \\ 
D & $8.23\pm0.03$ & $1.0\%$ & $2.15$ & $62\%$ & $0.15\pm0.02$ & $1.1\%$ & 0.79 \\ 
E & $8.34\pm0.03$ & $1.2\%$ & $2.56$ & $55\%$ & $0.45\pm0.03$ & $3.1\%$ & 1.40 \\ 
F & $8.77\pm0.10$ & $3.4\%$ & $4.28$ & $34\%$ & $1.11\pm0.06$ & $7.7\%$ & 1.45 \\
\hline                                
\end{tabular}
\tablefoot{The regions in the top panel of this table are defined as follows: 
    \tablefoottext{a}{The entire connected region, $3\sigma$ above the sky background.}
    \tablefoottext{b}{All overlapping sources masked out.}
    \tablefoottext{c}{As (b), but with the central seven bins masked out.}
    \tablefoottext{d}{The area within the central disc contour (see Fig.~\ref{fig:mass_map_radial_profile}a), with all masks applied.}
    \tablefoottext{e}{The area outside the central disc, with all masks applied.} \\
    The regions in the bottom panel are those defined in Section~\ref{sec:results_sfh}.
    The fractional masses and SFRs, \fracmass\ and \fracsfr, are defined relative to the Centre Masked region.
}
\end{table*}

\subsection{Star formation} \label{sec:results_ssfr}

\begin{figure}
    \centering
    \includegraphics[width=\hsize]{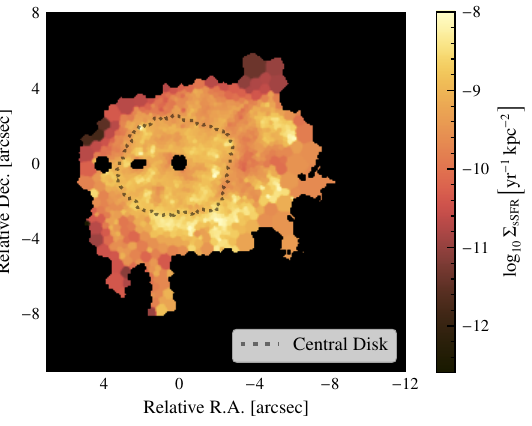}
    \caption{
        Spatially resolved map of the specific star formation rate density for F0083.
        As in Fig.~\ref{fig:mass_map_radial_profile}a, all external objects are masked out, and we display the central disc contour as a common point of reference.
    }
    \label{fig:F0083_ssfr_map}
\end{figure}

The entire galaxy shows a considerable amount of ongoing star formation, albeit concentrated primarily within the centre, and towards the direction of the tail in the SW quadrant.
In Fig.~\ref{fig:F0083_ssfr_map}, we show the spatially resolved specific star formation rate (\ssfr).
The \sfr\ derived in \textsc{bagpipes} is averaged over the last 100\,Myr, rather than the instantaneous measurement at time of observation (or the \sfr\ in the most recent age bin), so it is most comparable to the \sfr\ as traced by UV emission.
As shown in Table~\ref{tab:F0083_properties}, the integrated \sfr\ over the full region of F0083 is $20.8\pm2.6\,\sfrunits$, dropping to $14.9\pm0.9\,\sfrunits$ when the centre and other objects are masked.
Whilst significantly lower than the $\sfruv=26.2\pm0.9$ measured by \cite{rawle_star_2014}, we note that whilst their study considered the AGN contribution when deriving the \sfrir, they did not perform a similar analysis for \sfruv.
As discussed in Section~\ref{sec:method_contamination_masking}, this is therefore likely to be an overestimate of the UV contribution due to star formation alone.
Additionally, whilst our measurement may be a slight underestimate of the total \sfr\ due to the adopted masking procedure, the SED fitting method used here is able to account for dust attenuation, and so the measured \sfr\ is likely to be a closer reflection of the true value.

\subsection{Star formation histories} \label{sec:results_sfh}

\begin{figure*}
    \centering
    \includegraphics[width=\textwidth]{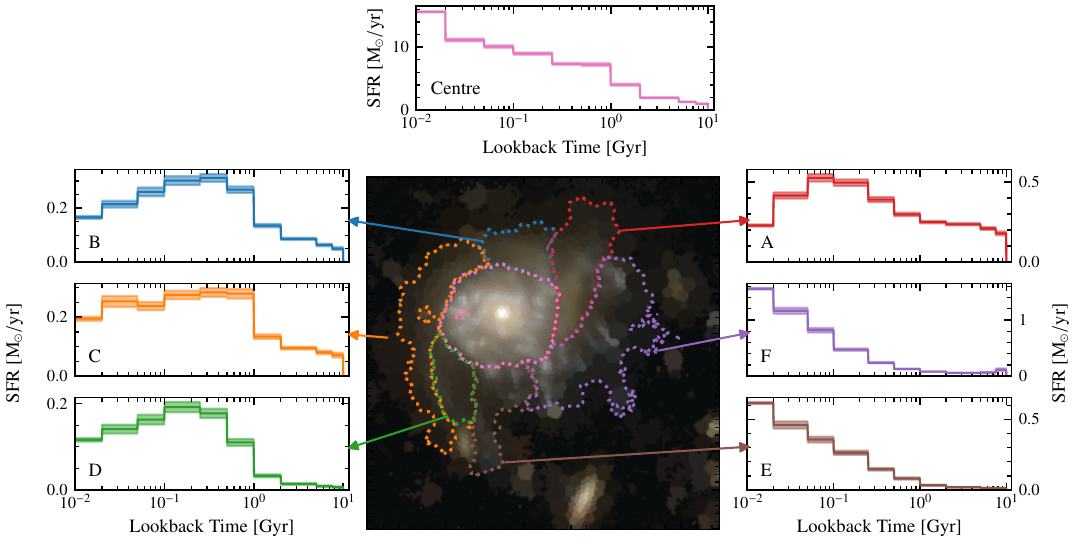}
    \caption{
        (Centre panel) RGB image of F0083, using the same filters as in Fig.~\ref{fig:F0083_full_colour_image}, constructed from the Voronoi-binned PSF-matched images.
        (Outer panels)  SFH in the central and exterior regions of the galaxy, derived as the sum of the individual SFHs of all bins in that region. 
        The region label (A-F) is inset.
        We note that the \sfr\ is plotted on a logarithmic scale, in contrast with Fig.~\ref{fig:bagpipes_example_fit}, and hence the area under the line no longer directly represents the total stellar mass formed in a given temporal range.
        We also note that whilst the youngest age bin spans $0\textrm{--}20\,$Myr, for clarity we have truncated the displayed range to only show $\tlb\geq10\,$Myr.
    }
    \label{fig:F0083_masked_regions_view_inset}
\end{figure*}

\begin{figure*}
    \centering
    \includegraphics[width=\textwidth]{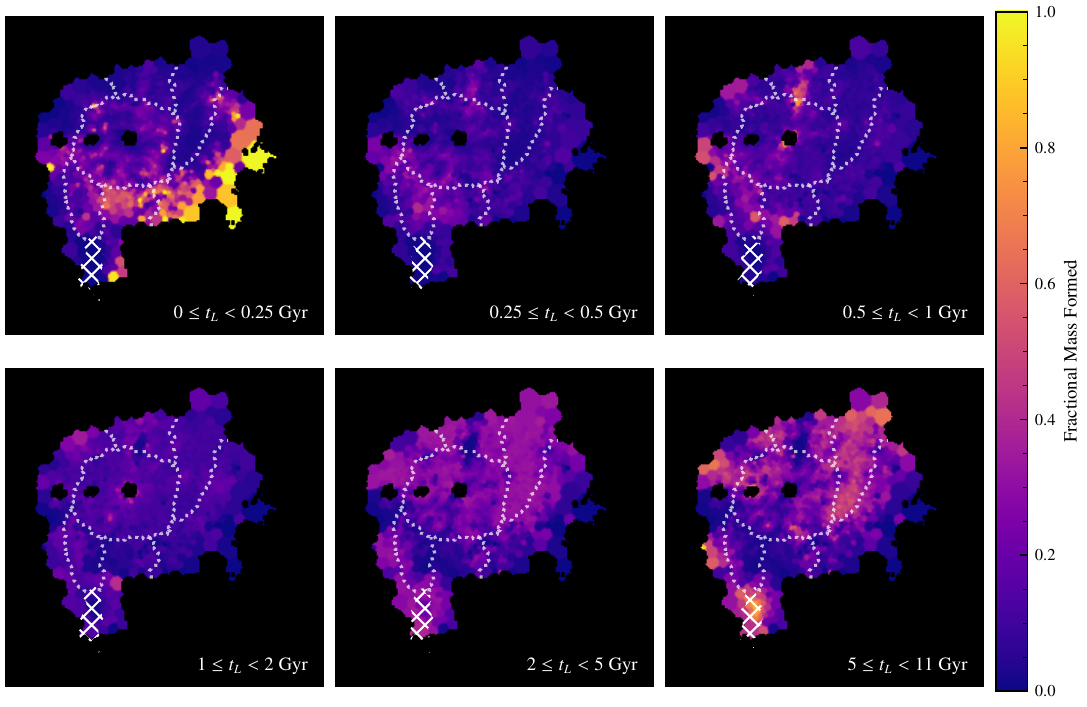}
    \caption{
        Fraction of the total stellar mass formed within different epochs, for each Voronoi bin. For ease of viewing, we have summed over the four most recent (0--0.25\,Gyr) and the two oldest (5--10\,Gyr) temporal bins.
        The dotted lines correspond to the different regions outlined in Fig.~\ref{fig:F0083_masked_regions_view_inset}.
        The cross-hatched area is a separate galaxy, at the same redshift as F0083 (see Section~\ref{sec:discuss_tidal_interaction}).
    }
    \label{fig:F0083_regions_age_maps}
\end{figure*}

There are multiple methods in which to display and analyse the resolved SFHs, and we focus here on two of the most pertinent.
In Fig.~\ref{fig:F0083_masked_regions_view_inset}, we decompose the exterior of F0083 (i.e. the area outside the central disc) into its constituent parts, based on a visual inspection of the colour image shown in Fig.~\ref{fig:F0083_full_colour_image}, forming six regions.
We distinguish between the two spiral arms (regions A and B), the putative tidal feature, and the blue knots and filaments comprising the tail.
The tidal feature is further separated into an inner and outer component (regions C and D), following the discontinuity in the colour image, and the tail divided into eastern and western regions (E and F), allowing for the possibility of different SFHs in the vicinity of the tidal feature.
We then calculate the integrated properties of these regions, by combining the results from all Voronoi bins within the contour, with the results displayed in Table~\ref{tab:F0083_properties}, and the SFHs in Fig.~\ref{fig:F0083_masked_regions_view_inset}.
We also calculate several additional quantities, used in previous studies (e.g. \citealt{werle_post-starburst_2022}) to compare the age and formation history of different stellar populations.
These include \mufrac, the fraction of the stellar mass which formed within the last 1.5\,Gyr, \agemw, the mass-weighted age, and \sfrratio, the ratio of the \sfr\ over the last 20\,Myr to the last 100\,Myr, all of which are tabulated in Table~\ref{tab:F0083_properties}.
We estimate \sfrratio\ to be accurate to $\approx20\%$ with the assumed priors and filter coverage available (see Appendix~\ref{app:sensitivity_sed_fitting} for further details), and comparisons of the different regions should be considered with this in mind.

We also look at the SFH of individual Voronoi bins.
We integrate the \sfr\ over a given temporal range to find the mass formed during that epoch, and compare this against the total mass formed since the beginning of the Universe.
The resulting maps of the mass fraction are shown in Fig.~\ref{fig:F0083_regions_age_maps}, with the previously defined regions overlaid as a consistent point of reference. 
We stress that there is a slight subtlety to these results.
These maps represent the fraction of the stellar mass formed at a given epoch, within the current bin location, and should not be confused with the distribution of the stellar mass at earlier times.
We summarise the salient points from these two figures below.

\begin{itemize}
    \item 
        Region A covers the majority of the north-western spiral arm, containing $\approx7\%$ of the total stellar mass in F0083.
        By any measure, this is the oldest region, with $\agemw=4.5$\,Gyr, and $\mufrac=20\%$.
        The inferred SFH in Fig.~\ref{fig:F0083_masked_regions_view_inset} shows a rising \sfr\ until $\approx0.1$\,Gyr ago, at which point it begins to decline.
        However, the overall variability in \sfr\ is low, with $\sfr_{\rm{max}}/\sfr_{\rm{min}}\approx2$.
        This is reflected in the fractional mass maps in Fig.~\ref{fig:F0083_regions_age_maps}, where the majority of the mass can be seen to have formed in the earliest time bins.
    \item
        Region B consists of the northern spiral arm.
        Overall, this area appears to have formed more recently than region A, with a lower \agemw\ and higher \mufrac, and a substantially lower \sfr\ in the earliest age bins.	
        Whilst this area has a currently declining \sfr, it appears to peak $\approx500$\,Myr ago, with a sudden increase at $\tlb\approx1$\,Gyr.
    \item 
        Region C covers the outermost part of the eastern tidal feature.
        The shape of the SFH is extremely similar to region B, with a sudden increase in \sfr\ at $\tlb\approx1$\,Gyr.
    \item
        Region D is the inner area of the eastern tidal feature.
        This area is much younger than the outer part, and contains just 1\% of the mass of F0083.
        Whilst the \sfr\ is extremely low at earlier epochs ($<0.02\,\sfrunits$ for $\tlb\geq2$\,Gyr), we see again the sharp increase at $\tlb\approx1$\,Gyr.
    \item 
        Regions E and F are what have been visually classified as the ``stripped tail'', with E indicating the area closest to the tidal feature (regions C and D).
        These regions account for $>10\%$ of the current \sfr\ of F0083, whilst containing $<5\%$ of the stellar mass.
        Both regions show a recently rising SFH with $\sfrratio>1$, and the resolved mass maps establish how recently the majority of mass has formed in these areas.
        We note that whilst region F shows a not insignificant amount of star formation in the earliest age bin, this can be attributed to the masking procedure.
        Considering Fig.~\ref{fig:F0083_regions_age_maps}, it is clear that our mask for region A has not fully encompassed the entirety of the old stellar population in the spiral arm -- or more accurately, the recently formed tail overlaps the existing stellar component along our line of sight.
    \item 
        The central disc itself contains a number of blue star-forming regions, and is where the majority of star formation is taking place.
        As with regions E and F, it has a recently rising SFH (see Table~\ref{tab:F0083_properties}).
        However, we note that there is a still faint imprint of the PSF snowflake pattern visible in the oldest age bin in Fig.~\ref{fig:F0083_regions_age_maps}, and so measurements in this region should be treated with more caution.
\end{itemize}

\section{Discussion} \label{sec:discussion}

As covered in the introduction, there are a variety of potential mechanisms that may act upon galaxies in a cluster environment.
Amongst these, we focus on tidal interactions with nearby galaxies, RPS due to infall into the cluster, and shock fronts from mergers of cluster substructures, and whether these physical mechanisms can explain the observed morphology and derived properties.

\subsection{Unwinding and tidal interactions} \label{sec:discuss_tidal_interaction}

One of the more unusual features of Figs.~\ref{fig:F0083_masked_regions_view_inset} and \ref{fig:F0083_regions_age_maps} is the extended south-eastern component (regions C and D), speculatively identified as a tidal feature in \cite{owers_shocking_2012}.
It is not immediately clear how this feature connects to other components of F0083 -- whether it is an extension of the northern spiral arm (region B), or its own distinct component.
High resolution JWST/NIRCam imaging favours the latter interpretation (see Fig.~\ref{fig:F0083_full_colour_image}), with the feature comprised of an outer redder portion (region C), similar to parts of the spiral arms, and an inner blue component (region D).
As noted in Section~\ref{sec:results_sfh}, the SFHs of these regions are unusual.
Region C consists of a great many older stars, and the mass formed at different epochs is consistent with the western and northern spiral arms (regions A and B).
However, approximately 1\,Gyr ago, regions B, C, and D all show a sharp increase in \sfr, peaking at $100\lesssim\tlb\lesssim500\,$Gyr. 

In \cite{bellhouse_gasp_2021}, the authors present evidence that in certain circumstances, RPS can cause an unwinding effect, whereby the spiral structure of the gas is preserved in the stripped material.
We do not find any particular evidence that this process is currently occurring in F0083, or is responsible for the observed structure evident in regions C and D.
The predicted and observed RPS-induced unwinding should primarily affect the distribution of gas in a stripped galaxy, and the resulting young stellar populations.
However, we see from Figs.~\ref{fig:mass_map_radial_profile}a and \ref{fig:F0083_regions_age_maps} that both the observed spiral structure and extended feature contain a considerable fraction of stars older than 1\,Gyr.
These components have therefore either formed in situ, or have been redistributed to their present locations after formation; neither are recent structures on the expected timescales of RPS ($\leq1\,$Gyr; \citealt{crowl_stellar_2008,lotz_gone_2019,owers_sami_2019,boselli_ram_2022}).

\begin{figure}
    \centering
    \includegraphics[width=\hsize]{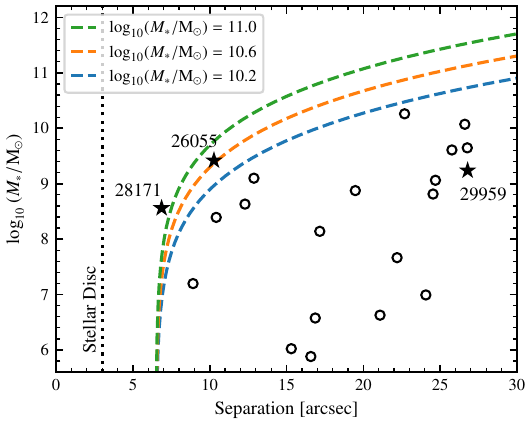}
    \caption{
        Potential interaction candidates for F0083, displayed in the projected distance-stellar mass plane as open circles.
        The approximate extent of the stellar disc is shown as a dotted black line, whilst we display the selection criterion for tidal interactions ($\atid/\agal=0.15$, see Eq.~\ref{eq:interaction_criterion}) as a series of dashed lines, covering a range of possible masses of F0083.
        The stars denote the objects under further study, with only two objects, 28171 and 26055, satisfying the interaction criterion.
    }
    \label{fig:tidal_interaction_selection}
\end{figure}

\begin{figure*}
    \centering
    \includegraphics[width=\textwidth]{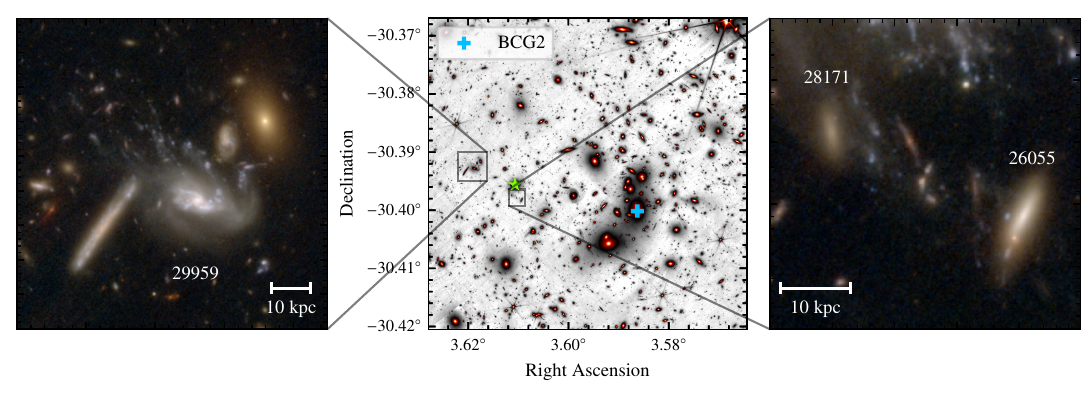}
    \caption{
        (Centre panel)  A2744 cluster field viewed through the NIRCam F200W filter, with the location of F0083 indicated by the green star, and BCG2 the blue cross.
        (Outer panels) RGB cutouts of the three tidal interaction candidates, using the same combination of filters and image scaling as in Fig.~\ref{fig:F0083_full_colour_image}.
       We note that the coverage of the cutout on the right overlaps with Fig.~\ref{fig:F0083_full_colour_image}.
    }
    \label{fig:F0083_neighbours_show_rgb}
\end{figure*}

We suggest instead that the observed mass asymmetry is more reminiscent of a tidal interaction with a companion object, such as observed in the Whirlpool galaxy (see Fig.~7 of \citealt{wei_spatially-resolved_2021} for an analogous plot).
This would also explain the observed increase in \sfr.
Given the apparent clockwise rotation of F0083, an object moving past the eastern side would interact with regions B, C and D in that order, in good agreement with the \tlb\ of the peak \sfr\ in each section (whether this is the only interaction, or simply the one which had the greatest effect on the \sfr, cannot be determined from the current data).
Under this assumption, it should be possible to locate the most likely interaction candidate from the surrounding field.
Tidal disturbances can be parameterised by the relative ratio of the radial tidal acceleration \atid, produced by the perturbating object, to \agal, the acceleration produced by the potential of the galaxy of interest.
Using the impulse approximation \citep{henriksen_tidal_1996,vollmer_new_2005}, these quantities can be expressed as 
\begin{align} \label{eq:henriksen_tidal_acceleration}
    \atid = {} & G \Mpert \left( \frac{1}{ (r-R)^2 } -\frac{1}{r^2} \right), \\
    \agal = {} & -\frac{G\Mgal}{R^2},
 \end{align}
where $r$ is the separation between the neighbour and galaxy of interest, \Mgal\ and \Mpert\ are the mass of the affected galaxy and the perturbating system respectively, $G$ is the gravitational constant, and $R$ is the distance from the centre of the affected galaxy to the outer edge of the tidal feature.
It should be noted that this equation differs slightly from that given in \cite{henriksen_tidal_1996}.
More precisely, the form given here corresponds to the tidal acceleration on the side of a galaxy closest to the perturbing mass, rather than the far side.
At $r\sim \mathcal{O}(R)$, as in this analysis, the tidal acceleration can no longer be assumed to be symmetric, and so this is the more appropriate form to use.
Following \cite{merluzzi_shapley_2016} and \cite{vulcani_gasp_2021}, we expect tidal interactions to have a significant influence on the morphology if the ratio
\begin{equation} \label{eq:interaction_criterion}
    \left\lvert\frac{\atid}{\agal}\right\rvert\gtrsim0.15.
\end{equation}
Thus, if the radial extent of the tidal stream $R$ is known, we can estimate the required \Mpert\ at a given distance that could be responsible.

\begin{figure}
    \centering
    \includegraphics[width=\columnwidth]{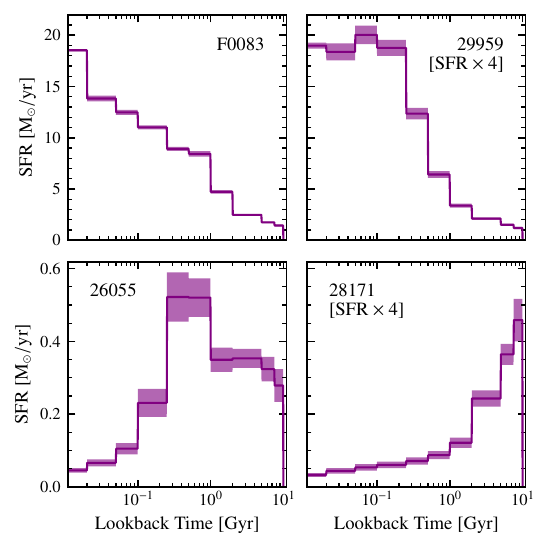}
    \caption{
        Star formation histories of the candidate galaxies for interactions with F0083, measured from the sum of the spatially resolved fits using \textsc{bagpipes} with the same method as in Section~\ref{sec:method_sed_fitting}.
        We note that the \sfr s for 29959 and 28171 have been multiplied by a factor of 4 in order to maintain the same $y$-axis scale.
    }
    \label{fig:neighbours_integrated_sfh}
\end{figure}

The exact ratio detailed in Eq.~\ref{eq:interaction_criterion} requires the total mass ratio and the true distance between the galaxies.
As neither of these are known, we use the stellar mass ratio and projected distance as proxies, within a narrow redshift range, noting that this can only give an approximate estimate of the tidal interaction.
We search for all objects with an on-sky separation of less than 30\arcsec\ ($\sim130\,\kpc$), and with redshifts $\lvert z-z_{\rm{F0083}}\rvert \leq 0.01$ ($\lvert\Delta\vpec\rvert\lesssim2300\,\kms$).
We use redshifts from the MegaScience catalogues, substituting the maximum-likelihood photometric redshifts with spectroscopic measurements where available, as detailed in Section~\ref{sec:method_redshift}.
The results are shown in Fig.~\ref{fig:tidal_interaction_selection}, with 30 objects satisfying these inital criteria.
We further refine our selection by estimating which objects satisfy Eq.~\ref{eq:interaction_criterion} for a radius $R=6\farcs5$ ($\sim30\,\kpc$), the maximum radial extent of the tidal feature (i.e. the outermost edge of region C).
As there is some uncertainty over the true stellar mass of F0083, we select all objects satisfying the equation over a range from $10.2\leq\log_{10}\left(M_*^{\rm{F0083}}/\rm{M_{\odot}}\right)\leq11.0$, shown as the dashed lines in Fig.~\ref{fig:tidal_interaction_selection}.
This leaves us with two interaction candidates, with UNCOVER IDs 28171 and 26055 (also known as ID 1272 in \citealt{watson_glass-jwst_2025}).
We investigate an additional galaxy (UNCOVER ID 29959) due to its highly disturbed morphology,\footnote{This galaxy is also one of the jellyfish candidates listed by \cite{rawle_star_2014}, named HLS001428-302334, although it was not fully covered by the HST/ACS imaging at the time.} despite not satisfying the tidal criterion at the present epoch.
We show the location and appearance of these objects in Fig.~\ref{fig:F0083_neighbours_show_rgb}, their integrated SFHs in Fig.~\ref{fig:neighbours_integrated_sfh}, and detail their properties below.
We note that there is an additional candidate galaxy, the ultra-diffuse galaxy $\approx8\,\arcsec$ west of F0083 (see Fig.~\ref{fig:F0083_full_colour_image}), previously identified by \cite{ikeda_near-infrared_2023}, with $\logstellmass\sim7.6$.
Due to the extremely faint photometry, and lack of spectroscopic information, we do not consider the derived redshift or stellar mass to be accurate enough to include this galaxy at present.

\subsubsection{Galaxy 29959} \label{sec:interaction_candidates_29959}

This galaxy is located 120\,\kpc\ north-east of F0083, at $\zspec=0.3019$ ($\deltavpecf=-320\,\kms$), with a mass $\logstellmass=9.42\pm0.07$.
It has previously been identified as a probable RPS galaxy by \cite{rawle_star_2014}, based on partial HST/ACS imaging, and named HLS001428-302334.
With the current JWST/NIRCAM imaging, it is clear that this galaxy has a highly disturbed morphology, with an extended tail of material stretching $\approx20\,\kpc$ towards the north-east, away from the cluster core.
The nearby edge-on disc, elliptical, and irregular galaxies are merely coincident on the sky, and are all located at substantially different redshifts ($\Delta\zphot>0.05$).
Due to the large spatial extent and asymmetry of this object, we opt to also display the full spatially resolved SFH in Appendix~\ref{app:spatially_resolved_29959}.

Based on the integrated SED fit shown in Fig.~\ref{fig:neighbours_integrated_sfh}, we find a continuously rising \sfr\ for 29959 until $\tlb\approx250\,$Myr, at which point it plateaus at $\sfr\approx4.5\,\sfrunits$.
Further detail is seen in the spatially resolved fit, with 29959 showing an enhanced \sfr\ in the tail and core.
A clear age gradient is visible, with the mass-weighted age increasing along the tail towards the stellar disc, and also from the core outwards to the edges of the galaxy, an indicator of outside-in quenching.
Taken together, these properties are highly suggestive of RPS of the gas content, as seen in many previous works \citep{gullieuszik_gasp_2017,bellhouse_gasp_2017,poggianti_gasp_2017, azevedo_spatially_2023}.
Regarding the interactions with F0083, the estimated tidal acceleration ratio for 29959 at the present separation is $\atid/\agal=6.5\times10^{-3}$.
It is therefore clear that 29959 cannot be responsible for any ongoing perturbation.
If F0083 had previously interacted with this galaxy, we might expect to see some discontinuity in the SFH, as gas-rich tidal interactions often trigger bursts of star formation \citep{woods_tidally_2006}.
However, there is no evidence supporting this scenario from our current observations, either in the integrated or spatially resolved SED fits.
As a result, we consider it extremely unlikely that the tidal feature in F0083 has any relation to 29959.

\subsubsection{Galaxy 28171}

This galaxy is embedded within the tidal feature of F0083 itself along our line of sight, at a projected distance of 30.8\,\kpc\ south-east of the core of F0083, and is also visible in both Figs.~\ref{fig:F0083_full_colour_image} and Fig.~\ref{fig:F0083_masked_regions_view_inset}.
No spectroscopic redshift is available, although the photometric redshift indicates a velocity offset of $\deltavpecf\approx-70\,\kms$, and we estimate the stellar mass to be $\logstellmass=8.56\pm0.06$.
The estimated tidal acceleration ratio for 28171 is $\atid/\agal=6.4$ (whilst this would seem to indicate that the tidal feature is gravitationally bound to 28171, we remind the reader that the projected distance $r$ likely underestimates the true 3D distance, particularly for photometric redshifts).
The integrated SFH presented in Fig.~\ref{fig:neighbours_integrated_sfh} shows a monotonically declining \sfr\ from the earliest age bin, although at a much lower level than the other galaxies considered.
Similarly, the spatially resolved stellar mass fraction shown in the hatched region of Fig.~\ref{fig:F0083_regions_age_maps} seems to indicate an inside-out quenching scenario, with the galaxy almost fully quiescent at the present epoch.
For both of these figures, we caution that the \sfr\ derived near the edges of 28171 may be driven by the unavoidable flux contamination from the tidal feature itself.
However, based on all of the available information, we cannot rule out a previous or ongoing tidal interaction with F0083.

\subsubsection{Galaxy 26055}

This galaxy is located at a projected distance of 46.1\,\kpc\ south-west of F0083, at $\zspec=0.306$ ($\deltavpecf=600\,\kms$), with an estimated mass $\logstellmass=9.24\pm0.09$. 
The mass estimate here should be regarded as a lower limit to the true mass, as $\approx20\%$ of the segmentation map for this object was masked out due to an overlap with a high-redshift interloper (ID 1256 at $z=2.7$, \citealt{watson_glass-jwst_2025}; visible in Fig.~\ref{fig:F0083_full_colour_image}).
We estimate the tidal acceleration ratio to be $\atid/\agal=0.25$ at the observed epoch.
From the SFH presented in Fig.~\ref{fig:neighbours_integrated_sfh}, we observe a relatively constant \sfr, with a burst occurring between $0.25\lesssim\tlb\lesssim1\,$Gyr, before declining to $\sfr\approx0.05\sfrunits$ at the present epoch.
This burst in the SFR of 26055 is a near perfect match to the age of the observed increase in SFR in the potential tidal regions of F0083, as discussed in Section~\ref{sec:results_sfh}.
We therefore consider it highly probable that 26055 interacted with F0083 at $0.5\lesssim\tlb\lesssim1\,$Gyr, triggering a burst in star formation in both galaxies.
The estimated tidal acceleration ratio indicates that 26055 may still be affecting the morphology of F0083, although as previously mentioned, this ratio may be overestimated due to projection effects.

\subsubsection{The cluster potential} \label{discuss:cluster_potential}

A similar analysis can also be performed to examine the tidal effects induced by the cluster potential, which has been shown to have a significant impact in some previous studies \citep{cortese_strong_2007}.
Using the best-fit values from \cite{ota_uniform_2004}, we estimate the total enclosed mass at the position of F0083 to be $\loghalomass=14.5$, at a separation of 580\,kpc (measured here from the X-ray peak; see Section~\ref{subsec:discuss_RPS} for further details). 
We estimate the total halo mass of F0083 to be $\loghalomass=11.9$ using the stellar mass-halo mass relation from \cite{behroozi_average_2013}, noting that the uncertainties on this relation are on the order of $\pm0.15\,$dex.
From this, we obtain a tidal acceleration ratio of $\atid/\agal\approx0.10$.
From the present observations, we therefore cannot rule out tidal interactions with the cluster potential as having had an effect on the morphology of F0083, although we expect this to be a substantially smaller effect than interactions with either of the neighbouring galaxies discussed earlier.

\subsection{Ram-pressure stripping} \label{subsec:discuss_RPS}

The most striking part of F0083 is the blue tail of stripped material, comprising many small knots, filaments and clumps stretching towards the south-west (regions E and F, Fig.~\ref{fig:F0083_masked_regions_view_inset}).
This has led to previous studies identifying F0083 as a potential jellyfish, or RPS, galaxy.
Using a simple analytic model, we can also assess the likelihood that F0083 is undergoing an RPS event, following a similar method as used elsewhere in the literature \citep[e.g.][]{cayatte_very_1994, bohringer_stripped_1997}, and the formalism of \cite{gullieuszik_gasp_2020}.
From \cite{gunn_infall_1972}, the ram pressure can be expressed as 
\begin{equation}
    P_{\rm{ram}} = \densICM(r)\times\Delta v^2,
\end{equation}
where $\densICM(r)$ is the ICM density at a clustercentric distance $r$, and $v$ is the velocity of the galaxy with respect to the cluster.
Abell 2744, however, has a highly asymmetric X-ray surface brightness distribution \citep{kempner_chandra_2004, owers_dissection_2011, chadayammuri_closing_2024}, and inferred gas density profiles can vary depending on the radial axis chosen. 
For this analysis, we obtain a first-order estimate using the work of \cite{ota_uniform_2004}.
They used ROSAT/HRI data to fit a $\beta$-model gas-density profile \citep{cavaliere_x-rays_1976} within the central 1.6\,Mpc, assuming isothermality (consistent with later Suzaka observations, see also \citealt{ibaraki_suzaku_2014}), parametrised as 
\begin{equation}
\densICM(r) = \densICM(0) \left(1+\frac{\rcl}{\rcore}\right)^{-3\beta/2},
\end{equation}
where the core radius $\rcore=132\farcs5$, the central gas density $\densICM(0)=8.08\times10^{-27}\cgsdens$, and $\beta=0.96$.
From this, we use the projected clustercentric distance (measured from the X-ray peak) to obtain $\densICM=3.09\times10^{-27}\cgsdens$ at the position of F0083.
Assuming a peculiar velocity offset of $\Delta\vpec=-712\,\kms$ (using the velocity estimate from Section~\ref{subsec:discuss_cluster_infall} does not substantially change our results), the estimated ram pressure is thus $P_{\rm{ram}}=1.6\times10^{-12}$\,N\,m$^{-2}$.
This is of a similar order to the estimated pressures for galaxies in the Virgo cluster \citep{chung_virgo_2007}, and an order of magnitude higher than other confirmed RPS systems \citep{gullieuszik_gasp_2017,fritz_gasp_2017}.

To test whether this is sufficient to strip the gas from F0083, we compute the anchoring force of the galaxy, defined as
\begin{equation}
\anchorforce=2\pi G\densGas\densStar,
\end{equation}
where \densGas\ and \densStar\ are the density profiles of the gaseous and stellar discs.
We express these as simple exponential discs,
\begin{equation}
    \Sigma=\left(\frac{M_{\rm{d}}}{2\pi r_{\rm{d}}^2}\right)e^{-r/r_{\rm{d}}},
\end{equation}
where $M_{\rm{d}}$ is the disc mass, $r_{\rm{d}}$ is the disc scale length, and $r$ is the radial distance from the centre of the galaxy.
We use our estimate of the total (measured) stellar mass for the stellar mass of the disc,\footnote{Using the extrapolated stellar mass (see Section~\ref{sec:results_mass}) changes the truncation radius by just $\approx1\%$.} $\mdstar=10^{10.24}\Msol$, and compute $\disclenstar=2.89\,\kpc$ following the relation of \cite{wu_scaling_2018}.
This is remarkably close to the expectation of $\disclenstar=r_{\rm{e}}/1.68=2.91\,\kpc$ (see Section~\ref{sec:results_mass}) for an exponential disc, justifying our model.
For the gas component, we assume a total mass $\mdgas=0.16\times\mdstar$ \citep{popping_evolution_2014}, and a scale length $\disclengas=1.7\times\disclenstar$ \citep{cayatte_very_1994}.
We can then compare the anchoring force at different radial distances from the centre of the galaxy.
The condition for stripping is met at the truncation radius \rtrunc, where $\anchorforce=P_{\rm{ram}}$.
At $r>\rtrunc$, the ram pressure exceeds the anchoring force, and the gas will be stripped.
From our model, we obtain an estimate of $\rtrunc\approx4.6\kpc$.
This corresponds to $\approx75\%$ of the initial gas mass that would have been stripped by ram pressure, consistent with the models of \cite{gullieuszik_gasp_2020}. 
This is likely to be a lower limit, as we have considered only the line-of-sight component of the velocity here (although \densICM\ may be similarly overestimated, as the projected distance will be less than the true distance to the cluster core).
From this, we conclude that it is highly likely that F0083 is indeed undergoing RPS.

Previous analyses have relied heavily on the HST/ACS imaging in the F435W, F606W, and F814W bands \citep{owers_shocking_2012,rawle_star_2014,lee_gmosifu_2022}.
Compared to these studies, there is a vast increase in the amount of high-resolution multi-wavelength data available (see Appendix~\ref{app:programmes} for further details).
One consequence is that with the expanded depth and wavelength coverage provided by JWST/NIRCam imaging, it is now clear that there is a significant spatial overlap between the south-western blue knots and filaments, and the underlying galactic disc.
This link is not visible in the earlier imaging -- \textit{c.f.} Fig.~\ref{fig:F0083_full_colour_image}, Fig.~2 of \cite{owers_shocking_2012}.
Whilst this does not change any of the overall conclusions regarding our model above, we posit that this may be due to ram pressure being exerted at a moderate angle with respect to the plane of the sky (see \citealt{roediger_ram_2006,akerman_how_2023} for examples of the variation in gas distribution due to inclination).
This inclination would account for the spectacular appearance of F0083, having both a considerable number of optically bright clumps, whilst these are only a short projected distance from the galactic disc.
In spectroscopic observations, this should be visible as an offset in the stellar and gas velocities.
Somewhat frustratingly, the existing GMOS IFU observations of \cite{lee_gmosifu_2022} have a S/N too low for reliable measurements of the stellar kinematics, and thus we conclude that this would provide an excellent opportunity for further research with deeper IFU observations.

Using the colours of the blue knots and filaments, \cite{owers_shocking_2012} estimated the upper limit on the age of these components to be $\approx100$\,Myr.
However, this is somewhat of a simplification, due to the overlap of the stellar component discussed above, and the assumption of a single stellar population.
From a visual inspection of the mass maps in Fig.~\ref{fig:F0083_regions_age_maps}, we see that a substantial number of Voronoi bins in regions E and F formed the majority of their mass within the last 100\,Myr.
Importantly, there is no other epoch during which a comparable amount of mass formed.
Additionally, looking at the integrated SFHs in Fig.~\ref{fig:F0083_masked_regions_view_inset}, both regions see a sharp increase in \sfr\ during the last 100\,Myr, with $\sfrratio>1$.
As such, we consider this result consistent with \cite{owers_shocking_2012}.
To summarise, within the last 100\,Myr, some mechanism has triggered a sudden increase in star formation in F0083, limited only to a single side of the galaxy.
There is considerable support for this being caused by RPS of the gas content, resulting from either the movement of this galaxy through the ICM, or the compression of the ICM by a shock front from mergers in the cluster itself.

\subsubsection{Shock fronts in the cluster environment} \label{subsec:discuss_cluster_env}

Considering the dynamics of the cluster environment discussed in Section~\ref{sec:intro}, it is unclear precisely which mechanism may have produced the stripped tail visible in F0083.
\cite{rawle_star_2014} suggest that the RPS of several galaxies, chiefly F0083 and 29959 as shown in Fig.~\ref{fig:F0083_neighbours_show_rgb}, is not driven by radial infall into the cluster, due to the tails being aligned away from the cluster core.
Instead, they may have been stripped by the passage of a shock front from mergers of the A2744 subclusters, which compressed the surrounding ICM.
Whilst initially plausible, there are several problems with this.
Firstly, for F0083, their suggested substructure passage is the interaction between the CTD and SMRC, using the terminology of \cite{owers_dissection_2011}, or the Main-1 and Main-2 haloes as described by \cite{furtak_uncovering_2023}.
However, as discussed in Section~\ref{subsec:discuss_RPS}, the largely unobscured view of the star-forming knots and filaments indicate that the direction of stripping is likely occurring out of the plane of the sky.
Taken with the orientation of the tail, for F0083 to be overrun with a shock front would require a merger in the NE-SW direction, towards the observer -- or if F0083 were part of the merging subcluster, the reverse orientation.
Neither of those scenarios has any clear observational support (see Section~\ref{sec:intro}), and F0083 has a projected offset of $>400\,\kpc$ from the closest known radio relic \citep{pearce_vla_2017,rajpurohit_dissecting_2021}.

Furthermore, we disagree with the characterisation of galaxy 29959 by \cite{rawle_star_2014}, who suggest that the morphology and tail orientation is strongly indicative of shock-front induced RPS, similar to F0083.
We concur that RPS is the most probable mechanism acting on this galaxy, and can rule out any ongoing interactions with F0083 (see Section~\ref{sec:interaction_candidates_29959}).
However, the more complete coverage of this object by JWST/NIRCam shows an extended star-forming tail pointing away from the cluster core, rather than towards the core as originally thought (we stress again that this was not possible to determine from the HST/ACS imaging available to \citealt{rawle_star_2014}).
In the simulations of \cite{yun_jellyfish_2019}, $\approx94\%$ of jellyfish galaxies have a tail anti-aligned to the direction of motion.
Whilst projection effects can have an impact on the observed tail orientation \citep{roediger_ram_2006}, the observed morphology thus indicates a non-zero velocity towards the cluster core.
Additionally, whilst there is little evidence here for the angle of stripping with respect to the line of sight, we suggest that the observed SFH in Section~\ref{sec:interaction_candidates_29959} and Appendix~\ref{app:spatially_resolved_29959} further supports this direction of stripping in the plane of the sky, in that the leading edge closest to the cluster core has a much higher \agemw\ than the opposing edge.
We conclude that there is little support from our observations of a coherent shock-front induced RPS, simultaneously affecting multiple galaxies in this region of A2744.

\subsubsection{Cluster infall} \label{subsec:discuss_cluster_infall}

\begin{figure}
    \centering
    \includegraphics[width=\columnwidth]{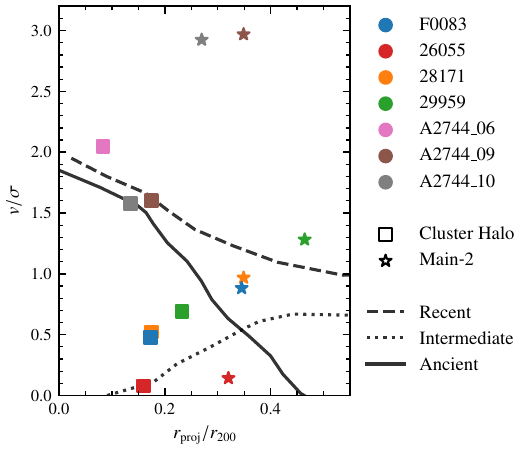}
    \caption{
    Projected phase space diagram for Abell 2744, showing the location of F0083, the tidal interaction candidates, and three reference jellyfish galaxies in the cluster \citep[A2744\_06, 09, and 10;][]{moretti_observing_2022}.
    The square points show the locations relative to the full cluster halo, with a virial radius of 2\,Mpc and $\sigma=1497\,\kms$, whereas the stars show the positions if we instead assume the Main-2 halo from \cite{furtak_uncovering_2023}, with $r_{200}=1\,$Mpc, and $\sigma=807\,\kms$ ($v/\sigma$ for A2744\_06  relative to Main-2 lies outside the plotted range).
    The coordinates of the halo centre are the same for both cases: $\rm{R.A.}=3.58641$, $\rm{Dec.}=-30.39997$.
    }
    \label{fig:pps_diagram}
\end{figure}

If we assume that the stripping of F0083 is solely caused by passage through the intracluster medium (ICM), it is easier to reconcile our observations with previous studies of the cluster substructure.
In Fig.~\ref{fig:pps_diagram}, we show the location of F0083 and the tidal interaction candidate galaxies on a projected phase-space (PPS) diagram.
We also include three known jellyfish galaxies from \cite{moretti_observing_2022} as a point of comparison (A2744\_06, 09, and 10), and delineate the approximate boundaries of the recent, intermediate, and ancient infalling populations, taken from \cite{rhee_phase-space_2017}.
The position of F0083 on the PPS would seem to indicate that it is already virialised, i.e. not undergoing first passage, although this makes assumptions about virialisation that do not necessarily hold true for a system as complex as A2744 \citep{rawle_star_2014,vulcani_early_2023}.
However, the position here within the PPS depends entirely on the expected membership of F0083 with respect to the cluster substructures. 
At a redshift $\zspec=0.3033$, F0083 is most consistent with being a member of the central substructures, likely the CTD using the formalism of \cite{owers_dissection_2011}.
If we eschew the overall cluster properties, and adopt instead the virial radius and velocity dispersion of the Main-2 halo of \cite{furtak_uncovering_2023}, the halo most consistent with the CTD substructure, F0083 lies closer to the edge of the recent ($<3.63\,$Gyr) infall region described by \cite{rhee_phase-space_2017}.
We stress that this does not mean that all galaxies in this region are definitely recent infallers, but that it is the most probable orbital solution given the observations.

If F0083 is only now stripped due to first infall, we can infer the most probable formation scenario.
The Main-2 halo contains BCG2, as indicated in Fig.~\ref{fig:F0083_neighbours_show_rgb}.
This is itself located at a redshift $\zspec=0.2997$, from which we infer a relative peculiar velocity for F0083 of $\vpec=830\,\kms$.
Without independent distance measurements, we cannot truly separate out the cosmological redshift from the Doppler redshift due to the peculiar velocity, although this estimate is consistent with the velocity dispersion of the halo as measured from strong lensing \citep{furtak_uncovering_2023}.
Indeed, if F0083 were located $1\,\Mpc$ in front of or behind BCG2 (the projected separation is just $360\,\kpc$), the inferred peculiar velocity would only range from $770\textrm{--}890\,\kms$.
We rule out F0083 being located at a higher cosmological redshift, with a negative peculiar velocity -- this is inconsistent with the observed morphology of the star-forming regions, as discussed in Section~\ref{subsec:discuss_RPS}.
We also stress that the orientation of the star-forming knots and filaments indicate a transverse velocity (\vtrans) component in the plane of the sky, directed away from the cluster core.
The most probable scenario is thus that F0083 is approaching (or has just passed) pericentre, on an orbit around the Main-2 halo, centred on BCG2.

Whilst both pre- and post-pericentre passages are consistent with the available data, previous studies have shown that most jellyfish galaxies in clusters are stripped on first infall, with a tail visible in the optical-NIR for up to $\sim600$\,Myr post-pericentre \citep{fumagalli_constraining_2011, yun_jellyfish_2019,roberts_lotss_2021,smith_new_2022}.
The tuned simulations of \cite{salinas_constraining_2024} also indicate that galaxies with the clearest signs of stripping typically have shorter durations of tail visibilities, and are generally high-speed first infallers.
This is certainly true for the most massive galaxy clusters, which contain few jellyfish galaxies with high \sfr s at small projected radii \citep{gullieuszik_gasp_2020}, and especially for those not undergoing edge-on stripping, the least efficient orientation for removing the gas content \citep{schulz_multi_2001,roediger_ram_2006,akerman_how_2023}.
We therefore submit that the most probable scenario is that F0083 is approaching pericentre on its first infall.

\subsection{Multiple physical processes} \label{sec:discuss_multiple_processes}

In Section~\ref{subsec:discuss_RPS} we establish that F0083 appears to be undergoing RPS, with a localised increase in \sfr\ at $\tlb\lesssim200\,$Myr.
Additionally, we show in Section~\ref{sec:discuss_tidal_interaction} that there is considerable evidence for a tidal interaction with a compact companion object, likely 26055, within the last $\approx1\,$Gyr.
As discussed in Section~\ref{sec:intro}, these represent clear examples of the two main subclasses of external physical processes expected to act on galaxies in dense environments.
From simulations, this is not necessarily an unexpected result -- \cite{marasco_environmental_2016}, looking at the HI content using the \textsc{eagle} simulation, found that the most common reason for gas removal was a combination of RPS and satellite interactions.\footnote{The term `tidal interactions' within their study refer to interactions with the group or cluster halo.}
However, this is the first analysis in which we are able to quantitatively estimate the contribution of these processes at $z\approx0.3$.
\cite{cortese_strong_2007}, analysing two galaxies at $z\approx0.2$, is the only other such analysis beyond those looking at low-redshift cluster galaxies \citep[e.g.][]{gavazzi_75_2001,vollmer_new_2005,scott_two_2012,fossati_muse_2019,serra_meerkat_2024}.
The authors there do not identify any potential interaction candidates, concluding that the combination of RPS and the tidal force from the cluster potential is responsible for the morphology and SFR of their galaxies.

The rarity of identifying both a galaxy-galaxy interaction and RPS in a single system stems primarily from the timescales on which these processes occur.
Within a cluster, the majority of interactions are likely to be high-speed interactions \citep[harassment,][]{moore_galaxy_1996}, and thus each individual interaction takes place over a very short period of time, consistent with the simulations of \cite{marasco_environmental_2016}.
By contrast, RPS may quench high-mass galaxies over timescales of up to 2\,Gyr \citep[][although the exact duration depends on a vast number of variables, such as the infall velocity, galaxy mass, and ICM density]{lotz_gone_2019}, however, the visible phase in the optical-NIR typically represents less than 600\,Myr post-pericentre \citep{roberts_lotss_2021,smith_new_2022,vulcani_relevance_2022,salinas_constraining_2024}.
The scenario proposed in this work, whereby F0083 may have undergone an interaction before the onset of RPS, has very rarely been observed amongst RPS galaxies at this or higher redshifts.

In previous studies of similar systems in the Local Universe, it has been proposed that these mechanisms are often linked, with the gravitational perturbations caused by a tidal encounter disrupting the potential well of a galaxy.
This would lower the anchoring force, and hence increase the efficiency of RPS \citep{gavazzi_75_2001,serra_meerkat_2024}.
Whilst we do not yet have sufficient constraints on the interaction history of F0083 to merit a more detailed model of this scenario, we consider it plausible, and hope that future spectroscopic analyses will be able to shed more light on the evolution of this system.

\section{Conclusion} \label{sec:conclusions}

We presented here the first observational analysis at $z>0.2$ quantitatively estimating the contribution of hydrodynamical and gravitational processes on a morphologically disturbed galaxy.
Using deep photometric data from GLASS, UNCOVER, MegaScience, and other publicly available surveys \citep[][see Appendix~\ref{app:programmes} for details]{weaver_uncover_2024, suess_medium_2024}, we derived the spatially resolved SFH of the galaxy F0083 using \textsc{bagpipes}, previously identified as a probable RPS system by \cite{owers_shocking_2012}, \cite{rawle_star_2014}, and \cite{lee_gmosifu_2022}.
F0083 is a high-mass ($\logstellmass = 10.24\pm 0.03$) spiral galaxy, with a tail of blue star-forming material extending towards the south-west from the stellar disc.
The resolved \sfr\ in this tail indicates an extremely high specific \sfr\ over the last 100\,Myr.
Additionally, the outer regions in all other quadrants show a decreasing \sfr\ at the observed epoch, with $\sfrratio<1$.
The only plausible conclusion is that F0083 is undergoing RPS, and we posit that the most probable scenario based on the available data is that F0083 is undergoing first infall, approaching pericentre.
We use the X-ray data of \cite{ota_uniform_2004} to construct a simple analytic model of RPS at the location of F0083, estimating that $\approx75\%$ of the initial gas mass has been stripped from the stellar disc.

F0083 also has an extended low surface brightness feature, stretching anticlockwise around the galaxy from the north to the south-east, reminiscent of the tidal features found in galaxies such as Messier 51.
In Fig.~\ref{fig:F0083_regions_age_maps}, we show that this feature is associated with an increased \sfr\ at $\tlb\lesssim1\,$Gyr, indicative of being formed by a previous interaction.
We verify that both this feature and the opposing spiral arms are present in the resolved stellar mass surface density map (Fig.~\ref{fig:mass_map_radial_profile}), and thus this unwinding effect cannot be attributed to RPS as in \cite{bellhouse_gasp_2021}.
Following the prescription of \cite{henriksen_tidal_1996}, we looked for all galaxies within a radius of 30\arcsec\ ($\sim140\,\kpc$) that satisfy Eq.~\ref{eq:interaction_criterion}, i.e. that are sufficiently massive at a close enough distance to exert a tidal force on F0083.
Despite the highly disturbed morphology, we find no evidence that the nearby galaxy 29959 (ID from UNCOVER DR3, also named HLS001428-302334 in \citealt{rawle_star_2014}) has previously interacted with or is sufficiently massive to currently exert a tidal force on F0083.
We can confirm that 29959 appears to be undergoing outside-in quenching as a result of RPS (see Appendix~\ref{app:spatially_resolved_29959}), and that given the orientation of the extended tail, it is highly likely that this object is on first infall towards the cluster centre.
As with F0083, we constructed a simple analytic model of RPS, estimating that 29959 has lost $\approx85\%$ of its initial gas mass.
We find that the nearby galaxy 26055 (also ID 1272 in \citealt{watson_glass-jwst_2025}) at $z_{\rm{grism}}=0.306$, separated from F0083 by a projected distance 46\,\kpc, is the most probable cause of the disturbed morphology, with its own SFH showing a sudden burst between $0.25\lesssim\tlb\lesssim1\,$Gyr.
We also identified a second interaction candidate, the quiescent elliptical galaxy 28171, embedded in the tidal stream along our line of sight.
Whilst we cannot rule out tidal interactions with the cluster potential as a possible explanation, at the present position of F0083, the tidal acceleration ratio from the cluster is substantially weaker compared to the estimated acceleration from either candidate galaxy.

We  demonstrated that with only deep photometric observations, it is possible to derive spatially resolved star formation histories that are sufficient to constrain multiple physical processes beyond the Local Universe.
We hope to confirm these results with follow-up observations, and apply the method more broadly in the future.

\begin{acknowledgements}
    We thank Alessandro Ignesti for his invaluable help in interpreting X-ray data, and the anonymous referee whose comments helped us to improve this paper.
    The data were obtained from the Mikulski Archive for Space Telescopes at the Space Telescope Science Institute, which is operated by the Association of Universities for Research in Astronomy, Inc., under NASA contract NAS 5-03127 for JWST. 
    These observations are associated with program JWST-ERS-1324. 
    We acknowledge financial support from NASA through grants JWST-ERS-1324.
    We also  acknowledge support from the INAF Large Grant 2022 ``Extragalactic Surveys with JWST'' (PI Pentericci). 
    B.V. and P.W are supported  by the European Union – NextGenerationEU RFF M4C2 1.1 PRIN 2022 project 2022ZSL4BL INSIGHT, and acknowledge support from the INAF Mini Grant ``1.05.24.07.01 RSN1: Spatially-Resolved Near-IR Emission of Intermediate-Redshift Jellyfish Galaxies'' (PI Watson).
    MT acknowledges support by the Australian Research Council Centre of Excellence for All Sky Astrophysics in 3 Dimensions (ASTRO 3D), through project number CE170100013. 
\end{acknowledgements}

\bibliographystyle{aa}
\bibliography{references} 

\begin{appendix}

\onecolumn
\section{Observing programmes} \label{app:programmes}

We show in Table~\ref{tab:obs_programmes} the details of all observing programmes which contributed to the MegaScience mosaics used in this study, including all observations within 30\arcsec\ of F0083.
For further details on the data reduction, we refer the reader to \cite{suess_medium_2024}.

\begin{table}[h]
\centering
\caption{All photometric observations used in this analysis.}\label{tab:obs_programmes}
\begin{tabular}{wc{2cm} wc{2cm} wc{1.75cm} wc{1.25cm} wr{2cm}}
\hline\hline
\multirow{2}{*}{Programme ID} & \multirow{2}{*}{PI} & \multirow{2}{*}{Instrument} & \multirow{2}{*}{Filter} & \multicolumn{1}{c}{Observation\small{$^{\strut}$}} \\
& & & & \multicolumn{1}{c}{Time (s)}\\
\hline
\multirow{3}{*}{HST-11689} & \multirow{3}{*}{R. Dupke} &  \multirow{3}{*}{ACS/WFC} & F435W$^{\strut}$ & 8081 \\ 
 & & & F606W & 6625 \\ 
 & & & F814W & 6624 \\ 
\midrule
\multirow{4}{*}{HST-13386} & \multirow{4}{*}{S. Rodney} & ACS/WFC & F814W & 446 \\ 
 & & \multirow{3}{*}{WFC3/IR} & F105W & 1915 \\
 & & & F125W & 806 \\ 
 & & & F160W & 1512 \\ 
\midrule
\multirow{2}{*}{HST-13389} & \multirow{2}{*}{B. Siana} & \multirow{2}{*}{WFC3/UVIS} & F275W & 22720 \\ 
 & & & F336W & 22720 \\ 
\midrule
\multirow{2}{*}{HST-13459} & \multirow{2}{*}{T. Treu} & \multirow{2}{*}{WFC3/IR} & F105W & 1068 \\ 
 & & & F140W & 1423 \\ 
\midrule
\multirow{3}{*}{HST-13495} & \multirow{3}{*}{J. Lotz} & \multirow{3}{*}{ACS/WFC} & F435W & 45747 \\ 
 & & & F606W & 20046 \\ 
 & & & F814W & 104270 \\ 
\midrule
\multirow{5}{*}{HST-15117} & \multirow{5}{*}{C. Steinhardt} & \multirow{2}{*}{ACS/WFC}& F606W & 1313 \\ 
 & & & F814W & 2357 \\ 
 & & \multirow{3}{*}{WFC3/IR} & F105W & 1412 \\ 
 & & & F125W & 1612 \\ 
 & & & F160W & 1612 \\ 
\midrule
\multirow{1}{*}{HST-15940} & \multirow{1}{*}{B. Ribeiro} & WFC3/UVIS & F225W & 22674 \\ 
\midrule
\multirow{3}{*}{JWST-1324} & \multirow{3}{*}{T. Treu} & \multirow{3}{*}{NIRISS} & F115WN & 5497 \\ 
 & & & F150WN & 5497 \\ 
 & & & F200WN & 2749 \\ 
\midrule
\multirow{7}{*}{JWST-2561} & \multirow{7}{*}{I. Labbe} & \multirow{7}{*}{NIRCam} & F115W & 18940 \\ 
 & & & F150W & 18940 \\ 
 & & & F200W & 11725 \\ 
 & & & F277W & 13399 \\ 
 & & & F356W & 13399 \\ 
 & & & F410M & 13399 \\ 
 & & & F444W & 16492 \\ 
\midrule
\multirow{6}{*}{JWST-2756} & \multirow{6}{*}{W. Chen} & \multirow{6}{*}{NIRCam} & F115W & 4209 \\ 
 & & & F150W & 4209 \\ 
 & & & F200W & 4209 \\ 
 & & & F277W & 4209 \\ 
 & & & F356W & 4209 \\ 
 & & & F444W & 4209 \\ 
\midrule
\multirow{4}{*}{JWST-2883} & \multirow{4}{*}{F. Sun} & \multirow{4}{*}{NIRCam} & F182M & 5154 \\ 
 & & & F210M & 8761 \\ 
 & & & F360M & 2061 \\ 
 & & & F480M & 2577 \\ 
\midrule
\multirow{3}{*}{JWST-3516} & \multirow{3}{*}{J. Matthee} & \multirow{3}{*}{NIRCam} & F070W & 13915 \\ 
 & & & F090W & 17598 \\ 
 & & & F356W & 5261 \\ 
\midrule
\multirow{5}{*}{JWST-3538} & \multirow{5}{*}{E. Iani} & \multirow{5}{*}{NIRCam} & F210M & 258 \\ 
 & & & F300M & 773 \\ 
 & & & F335M & 515 \\ 
 & & & F410M & 773 \\ 
 & & & F460M & 1288 \\ 
\midrule
\end{tabular}
\end{table}

\twocolumn

\begin{multicols}{2}[{
\begin{table*}
\centering
    \setcounter{table}{0}
    \caption{(\textit{continued}) All photometric observations used in this analysis.}\label{tab:obs_programmes2}
    \begin{tabular}{wc{2cm} wc{2cm} wc{1.75cm} wc{1.25cm} wr{2cm}}
    \hline\hline
    \multirow{2}{*}{Programme ID} & \multirow{2}{*}{PI} & \multirow{2}{*}{Instrument} & \multirow{2}{*}{Filter} & \multicolumn{1}{c}{Observation\small{$^{\strut}$}} \\
    & & & & \multicolumn{1}{c}{Time (s)}\\
    \hline
    \multirow{13}{*}{JWST-4111} & \multirow{13}{*}{K. Suess} & \multirow{13}{*}{NIRCam} & F070W$^{\strut}$ & 7215 \\ 
     & & & F090W & 14430 \\ 
     & & & F140M & 7215 \\ 
     & & & F162M & 7215 \\ 
     & & & F182M & 7215 \\ 
     & & & F210M & 7215 \\ 
     & & & F250M & 8246 \\ 
     & & & F300M & 8246 \\ 
     & & & F335M & 8246 \\ 
     & & & F360M & 8246 \\ 
     & & & F430M & 8246 \\ 
     & & & F460M & 8246 \\ 
     & & & F480M & 8246 \\
    \hline
    \end{tabular}
    \tablefoot{Filters ending in ``N'' denote those used in NIRISS, to avoid confusion with the NIRCam filters of the same name.}
\end{table*}
}]
\end{multicols}

\flushqueue

\section{PSF matching} \label{app:psf_matching}

Three of the filters to be matched to F444W are the F115W, F150W and F200W filters on JWST/NIRISS.
Whilst these are in general very similar to the corresponding filters on JWST/NIRCam, we derive separate PSFs to account for the different native pixel scales of the detectors, and the pixelation effects when redrizzled to the mosaic pixel scale of $0\farcs04/$pix.
For these, we directly follow the method used by \cite{weaver_uncover_2024}, and select point sources based on their position in the size-magnitude plane, as traced by the ratio of fluxes within 0\farcs16 and 0\farcs32 diameter apertures, against the magnitude within the 0\farcs32 aperture.
We use the default parameters of \textsc{aperpy} \citep{weaver_uncover_2024} to generate a field-averaged PSF for each filter, rejecting sources with an AB magnitude fainter than 24, with nearby contamination, or with a cutout signal/noise (S/N) below 800.
For completeness, we also derive PSFs for the F090W and F158M filters, even though the coverage in these bands does not overlap with any of the galaxies we are interested in.

For the WFC3/UVIS filters F225W, F275W, and F336W, there are no suitable unsaturated, uncontaminated point sources available in the A2744 mosaic.
We therefore derive a PSF based on archival imaging.
There exists already an empirical PSF model, in the STDPSF format \citep{anderson_empirical_2016}; however, this covers only the central $25\times25$ pixels, and not the extended diffraction spikes.
Similarly, the WFPC3 PSF database \citep{dauphin_wfpc2_2021} offers cutouts, but extending only to a $51\times51$ pixel area, significantly smaller than the $101\times101$ pixel cutouts and kernels derived in \cite{suess_medium_2024}.
We therefore use the PSF database to extract larger cutouts from the original images.
For each filter, we select all PSFs matching the following criteria (modified from the ``Good Quality PSF Subset" selection):
\begin{itemize}
    \item A telescope focus between $-2$ and 2 $\mu$m.
    \item The PSF $x$-centre and $y$-centre falling between 256 and 3835 (i.e. not within 256 pixels of the exterior edges of the detector).
    \item A PSF flux between 20,000 and 90,000 electrons.
    \item A PSF fit quality parameter (\textsc{qfit}) less than 0.1.
    \item An exposure time less than 900 seconds.
    \item An observation date after 9th September 2014.
\end{itemize}

\begin{figure}
    \centering
    \vspace{8.25cm}
    \includegraphics[width=\columnwidth]{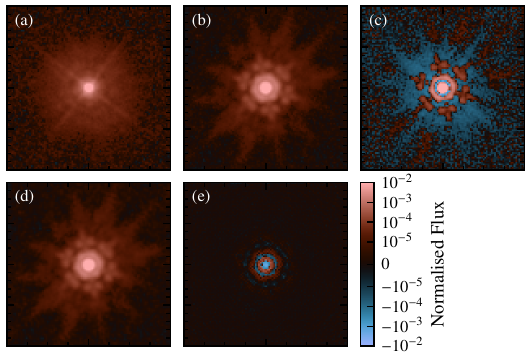}
    \caption{
    Visual representation of the PSF-matching procedure. All cutouts are 2\arcsec\ per side.
    (a) Empirical PSF for the F275W filter in the mosaic frame, with the total flux normalised to unity.
    (b) Target PSF, i.e. the normalised F444W PSF from the MegaScience data release.
    (c) Convolution kernel as derived by \textsc{pypher}.
    (d) Original PSF (a) convolved with the kernel (c).
    (e) Residuals of the convolved PSF compared to the target, i.e. $\text{(d)}-\text{(b)}$.
    }
    \label{fig:psf_kernel_comparison}
\end{figure}

Using this selection, we cutout $201\times201$ pixel regions from the original \texttt{*.flc} images centred on each PSF, subtracting the \textsc{mdrizsky} value present in the FITS header, and apply the same contamination rejection procedure as for the NIRISS PSFs.
We relax the constraints on the S/N to allow any cutout with an integrated $\text{S/N}>200$, and an AB magnitude greater than 26.
This leaves us with $\approx200$ sources for each filter, which we stack using \textsc{aperpy} to produce an empirical detector-frame PSF.
This is then reprojected to the WCS of the MegaScience mosaics, accounting for the detector distortion and orientation, and cropped to $101\times101$ pixels to match the existing PSFs.

Convolution kernels for the PSFs are derived using \textsc{pypher} \citep{boucaud_convolution_2016}, which computes kernels using an algorithm based on Wiener filtering, without any assumptions on the spherical symmetry of the system.
We set the regularisation parameter to $3\times10^{-3}$, to limit high-frequency noise, and oversample each PSF by $3\times$ before matching, using a 3rd order spline interpolation.
In Fig.~\ref{fig:psf_kernel_comparison}a, we show an example of the original PSF for the F275W filter of WFC3/UVIS, compared against the target PSF in  Fig.~\ref{fig:psf_kernel_comparison}b.
The differences between the diffraction spikes, caused by the secondary mirror support structures, are clearly visible.
In Fig.~\ref{fig:psf_kernel_comparison}c, we show the \textsc{pypher}-derived convolution kernel, with the convolved PSF in Fig.~\ref{fig:psf_kernel_comparison}d, and the residuals against the target in Fig.~\ref{fig:psf_kernel_comparison}e.
There is still some structure visible in the residuals, most prominently in the very centre of the PSF.
Whilst this fit can be improved by relaxing the regularisation parameter in \textsc{pypher}, doing so will also increase higher-frequency noise across the image.

\FloatBarrier

\section{Photometric uncertainties} \label{app:photometric_uncertainties}

\begin{figure}[h]
    \centering
    \includegraphics[width=\columnwidth]{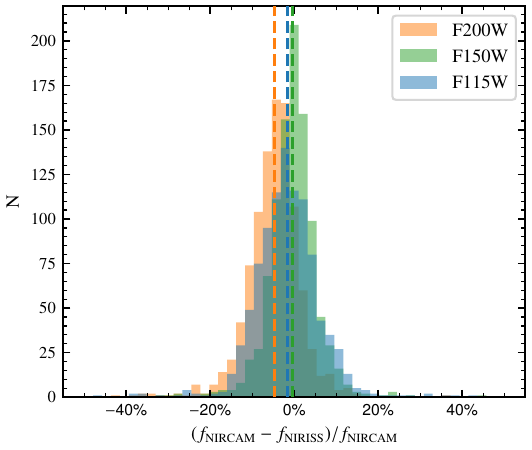}
    \caption{
    Relative offset between the measured fluxes in NIRCam and NIRISS in all Voronoi bins for F0083.
    The dashed lines show the median of the distribution for each of the three filters.
    }
    \label{fig:photometric_uncertainties}
\end{figure}

With the exception of F158M, all photometric filters on NIRISS originated as ``flight spares'' for NIRCam.
As such, these filters are extremely similar to their NIRCam counterparts, and we exploit this to provide an estimate of the systematic uncertainty in the photometric measurements (not to be confused with the random error; see Section~\ref{sec:method_voronoi_binning}).
At the position of F0083, only three filters have coverage in both NIRCam and NIRISS: F115W, F150W, and F200W.
In Fig.~\ref{fig:photometric_uncertainties}, we show the relative offset between the fluxes in each instrument, for each of the three filters, as measured after convolving to a common PSF and Voronoi-binning to a S/N of 100 (see Sections~\ref{sec:method_psf_matching} and \ref{sec:method_voronoi_binning}).
Typical measured uncertainties are 1.7\%, 1.6\% and 2.0\% for F115W, F150W, and F200W respectively.
We find the best agreement for F150W, with a median offset of 0.0\%, and $\sigma=4.6\%$.
F115W shows the largest spread, with $\sigma=6.3\%$, and a small offset of $-1.5\%$ (i.e. the measured flux is greater in NIRISS than NIRCam).
F200W has a standard deviation of 5.1\%, but a clear systematic offset of $-4.3\%$.
As such, we consider an additional uncertainty of 5\% sufficient to account for these systematic variations.

\section{Sensitivity of the SED fitting} \label{app:sensitivity_sed_fitting}

\begin{figure}
    \centering
    \includegraphics[width=\columnwidth]{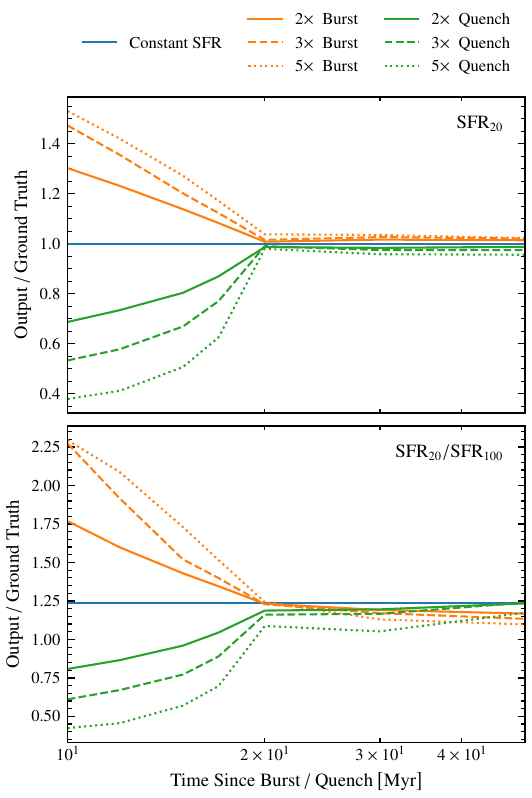}
    \caption{
    Ratio of the estimated value from SED fitting to the ground truth as used in the input models.
    This is shown as a function of the age of the instantaneous change in \sfr, for a range of bursting and quenching scenarios.
    (Top) Reliability of the \sfr\ measured over the previous 20\,Myr.
    (Bottom) Accuracy of the ratio of the \sfr\ measured over 20\,Myr to the last 100\,Myr.
    }
    \label{fig:plot_sfrratio_model}
\end{figure}

The overall accuracy of SED fitting in retrieving the SFH is a well-studied topic, and the use of non-parametric priors has been covered extensively in previous works \citep[e.g.][]{leja_how_2019,lower_how_2020,tacchella_fast_2022,suess_recovering_2022,pacifici_art_2023,ciesla_goods-alma_2023,wan_stochastic_2024,wang_population_2025}.
Here, we focus only on how well-constrained the SFR\ is on very short timescales, with our choice of priors and the available rest-frame wavelength coverage.
We construct a simple SFH, consisting of a constant SFR beginning at $\tlb=100\,$Myr, and instantaneously increasing or decreasing by a constant factor at $10\leq\tlb\leq50\,$Myr.
We apply this SFH to a set of model galaxies constructed using \textsc{bagpipes}, covering a range of metallicities, dust extinctions, and ionisation parameters.
We then construct mock SEDs using these model galaxies, assuming the same redshift ($z=0.3$) and filter coverage as used in this analysis.
The S/N of each filter in this mock catalogue is taken from the median S/N in each filter for the Voronoi-binned observations of F0083.
We fit this catalogue using \textsc{bagpipes} with the same priors as described in Section~\ref{sec:method_sed_fitting}, and show the retrieved $\sfr_{20}$ and \sfrratio\ in Fig.~\ref{fig:plot_sfrratio_model}.

From this, several results are immediately apparent.
Firstly, although the absolute accuracy of $\sfr_{20}$ for abrupt changes below 20\,Myr is low, the combination of the chosen prior and the rest-frame filter coverage is still highly sensitive to the \sfr\ on these short timescales.
In general, the fitted $\sfr_{20}$ overestimates the change in \sfr, predicting increasingly stronger bursts or quenching than the model the more recently the change occurred.
That all quenching, bursting, and constant scenarios can retrieve the correct value of $\sfr_{20}$ when the change in \sfr\ occurs at $\tlb=20\,$Myr indicates that this is likely an artefact caused by the width of the age bins.
Similar results were found by \cite{ciesla_goods-alma_2023}, who found that the accuracy of the derived \sfr\ decreased rapidly at $\tlb<30\,$Myr, the edge of their most recent age bin.

Secondly, even when we assume a constant \sfr, our derived values of \sfrratio\ are consistently biased slightly higher (for bursts/quenching occurring at $\tlb\geq20\,$Myr).
We thus cannot distinguish between a recent burst with a constant \sfr, and one with a slightly rising \sfr, where the difference in \sfrratio\ is $\approx20\%$.
We take this as an indication of the uncertainty on \sfrratio.
We do however note that in all quenching scenarios, we still correctly estimate the direction of change of \sfr, i.e. \sfrratio\ is always less than unity.
This positive bias can be intuitively understood by considering the nature of the continuity SFH prior, which disfavours rapid changes in the SFR between adjacent age bins \citep{leja_how_2019}.
Rather than deriving a sudden rise in SFR at $\tlb=100\,$Myr, as in our model SFH, the posterior distribution includes a small rise in SFR within the previous age bin ($100<\tlb<250$\,Myr), lowering the derived value of $\sfr_{100}$.

Although this somewhat contrived scenario may not necessarily reflect a realistic SFH, we consider it a useful exercise to estimate the accuracy of our results.
Our conclusions are that:
\begin{itemize}
    \item We cannot accurately derive abrupt changes to the SFR on timescales less than 20\,Myr.
    \item We can reliably retrieve the direction of change of SFR on timescales below 100\,Myr, to an overall accuracy of $\sim20\%$. 
\end{itemize}

\section{Spatially resolved properties of galaxy 29959} \label{app:spatially_resolved_29959}

\begin{figure}
    \centering
    \includegraphics[width=\columnwidth]{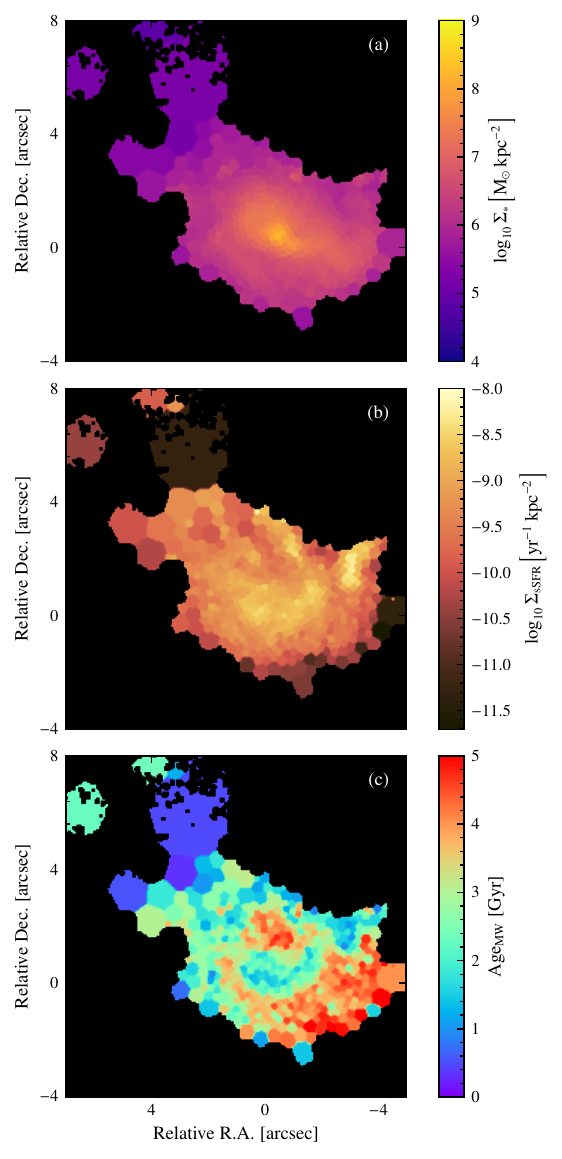}
    \caption{
    (a) Spatially resolved map of the stellar mass surface density for 29959, with all external objects masked out.
    The discontinuity in the tail is due to the very low S/N in this area.
    (b) Specific star formation rate density. 
    We note that the lower limit of the colour map differs from Fig.~\ref{fig:F0083_ssfr_map}, in order to preserve as much detail as possible.
    (c) Mass-weighted age, \agemw, expressed on a linear scale in units of Gyr.
    }
    \label{fig:A2744_COMPANION_regions_MW_age_SFR}
\end{figure}

We present in Fig.~\ref{fig:A2744_COMPANION_regions_MW_age_SFR} some of the spatially resolved properties of the galaxy 29959.
The binning procedure and SED fitting were carried out in the same manner as for F0083 in the main analysis.
This leads to a small discontinuity in the tail in the maps of 29959, as the S/N in this location in the F150W filter was too low to isolate the object of interest.
As discussed in Section~\ref{sec:interaction_candidates_29959}, 29959 shows clear-signs of outside-in quenching, with the centre still actively star-forming in Fig.~\ref{fig:A2744_COMPANION_regions_MW_age_SFR}b, and \agemw\ increasing with increasing radius.
This latter effect is not symmetric however -- the leading edge of 29959 (closest to the cluster centre) is substantially older than the trailing edge, particularly the extended tail.
We posit that this age gradient is a clear indicator of RPS as seen in many previous studies \citep{poggianti_gasp_2017,azevedo_spatially_2023}.
Performing a similar analysis as in Section~\ref{subsec:discuss_RPS}, with a peculiar velocity offset of $\Delta\vpec=1030\,\kms$, we can estimate the ram pressure exerted on 29959 as $P_{\rm{ram}}=2.61\times10^{-12}$\,N\,m$^{-2}$.
Assuming the same scaling relations as previously, and exponential stellar and gas discs, we obtain $\rtrunc\approx1.7\kpc$, which corresponds to $\approx85\%$ of the initial gas mass lost.
It is therefore highly likely that 29959 is undergoing RPS, although dedicated spatially resolved spectroscopy would be required to definitively rule out previous or ongoing interactions.

\end{appendix}
  
\end{document}